\documentclass[AMA,LATO1COL]{WileyNJD-v2}

\usepackage{amssymb}
\usepackage{subfig}
\usepackage{placeins}

\articletype{Research Article}%

\received{}
\revised{}
\accepted{}

\begin{document}

\title{Preventing wind turbine tower natural frequency excitation with a quasi-LPV model predictive control scheme\protect\thanks{\textbf{Short title}: A qLPV-MPC framework for preventing tower natural frequency excitation}}

\author[1]{Sebastiaan Paul Mulders}
\author[2]{Tobias Gybel Hovgaard}
\author[2]{Jacob Deleuran Grunnet}
\author[1]{Jan-Willem van Wingerden}

\authormark{S.P. MULDERS \textsc{et al}}

\address[1]{\orgdiv{Delft Center for Systems and Control, Faculty of Mechanical Engineering}, \orgname{Delft University of Technology}, \orgaddress{\country{The Netherlands}}}
\address[2]{\orgdiv{Vestas Power Solutions}, \orgname{Vestas Wind Systems A/S}, \orgaddress{\country{Denmark}}}

\corres{Sebastiaan Paul Mulders, Delft Center for Systems and Control, Faculty of Mechanical Engineering, Mekelweg 2, 2628 CD Delft, The Netherlands.\\ \email{s.p.mulders@tudelft.nl}}

\abstract[Abstract]{With the ever increasing power rates of wind turbines, more advanced control techniques are needed to facilitate tall towers that are low-weight and cost effective, but in effect more flexible. Such soft-soft tower configurations generally have their fundamental side-side frequency in the below-rated operational domain. Because the turbine rotor practically has or develops a mass imbalance over time, a periodic and rotor-speed dependent side-side excitation is present during below-rated operation. Persistent operation at the coinciding tower and rotational frequency degrades the expected structural life span. To reduce this effect, earlier work has shown the effectiveness of active tower damping control strategies using collective pitch control. A more passive approach is frequency skipping by inclusion of speed exclusion zones, which avoids prolonged operation near the critical frequency. However, neither of the methods incorporate a convenient way of performing a trade-off between energy maximization and fatigue load minimization. Therefore, this paper introduces a quasi\nobreakdash-linear parameter varying model predictive control (qLPV\nobreakdash-MPC) scheme, exploiting the beneficial (convex) properties of a qLPV system description. The qLPV model is obtained by a demodulation transformation, and is subsequently augmented with a simple wind turbine model. Results show the effectiveness of the algorithm in synthetic and realistic simulations using the NREL~5\nobreakdash-MW reference wind turbine in high-fidelity simulation code. Prolonged rotor speed operation at the tower side-side natural frequency is prevented, whereas when the trade-off is in favor of energy production, the algorithm decides to rapidly pass over the natural frequency to attain higher rotor speeds and power productions.}


\maketitle

\section{Introduction}\label{sec:I}
The tower makes up a substantial part of the total turbine capital costs, and therefore finding an optimum between its mass and manufacturing expenses is a critical trade-off.\cite{ref:Dykes2018TallTowers} For conventional towers, diameters are limited because of land-based transportation constraints. This aspect dictates the increase of wall thickness for the production of taller towers, and consequently leads to increased weight and costs. Conventional tower designs are soft-stiff to locate the tower fundamental frequency outside the turbine variable-speed operational range, and thereby eliminate the possibility of exciting a tower resonance by the rotor rotational or blade-passing frequency. However, with the ever increasing wind turbine power rates, a combination of technical solutions should enable future, low-cost, tall towers, by relaxing this frequency constraint. Soft-soft tower configurations form an opportunity for tall towers by their smaller tower diameters and reduced wall thickness. As a result, soft-soft towers are less stiff, and have their natural frequency in the turbine operational range. Therefore more advanced control solutions will be key in avoiding the excitation of these frequencies for extended periods of time. 
 
In practical scenarios, the center of mass of the wind turbine rotor assembly does not coincide with the actual rotor center as a result of, e.g., manufacturing imperfections, wear and tear, fouling, and icing.\cite{ref:Hau2013WTBook} Moreover, vibrations are also induced by rotor aerodynamic imbalances caused by pitch errors and damage to the blade surface.\cite{ref:GL2012OffshoreGuidelines} Consequently, during variable-speed below-rated operation, the rotor rotational or blade-passing frequency may excite the structural side-side natural frequency. Small perturbations can lead to load fluctuations comparable to fore-aft stresses, because the turbine rotor provides negligible aerodynamic side-side damping, at an order of magnitude smaller than the fore-aft damping ratio.\cite{ref:Burton2001WindEnergyHandbook} As a result, excitation of the side-side mode possibly results in accelerated and accumulative fatigue damage. 

Straightforward control implementations are available for reducing and mitigating the excitation of the tower fore-aft and side-side modes. An active method for reducing tower motion is the use of an integrated nacelle acceleration signal in a proportional feedback structure. Depending on the measured acceleration direction, the resulting signals form an addition to the collective pitch\cite{ref:Bossanyi2003LoadReductions, ref:Bossanyi2012FieldCART3} or generator torque\cite{ref:Wright2011RefinementsCART} control signal, for respective damping of fore-aft and side-side vibrations. Another, more passive method, entails the prevention of structural mode excitation by manipulating the generator torque when the rotor speed approaches the excitation frequency.\cite{ref:Bossanyi200DesignFeedback} This method is often referred to as \textit{frequency skipping} by inclusion of \textit{speed exclusion zones}.

All of the above described active and passive methods complicate the controller design, by requiring extra proportional-integral-derivative (PID) feedback control loops with additional logic and speed set-points. Also, the methods do not incorporate convenient and inherent tuning capabilities for a trade-off between produced energy and fatigue loading. Therefore, more advanced control algorithms might form a solution by providing a more integrated way of controller synthesis, incorporating power and load objectives. While an abundance of publications on advanced wind turbine control algorithms outline the possible benefits\cite{ref:Bianchi2007WTCS}, to the authors' knowledge, more sophisticated control methods do not yet see a wide-spread adoption in industrial-grade wind turbine control systems; PID control structures\cite{ref:Mulders2018DRC} provide ease of implementation while resulting in a sufficient performance level.

An advanced control method, that has seen a substantial gain of interest from industry in the past decades, is model predictive control (MPC).\cite{ref:Rawlings2009BookMPC, ref:Hovgaard2015WindPowerGradients} The most evident benefits of MPC over PID control\cite{ref:Holkar2010OverviewMPC} are (1) the ability of including constraints, (2) coping with the complexity of nonminimum phase systems, (3) robustness against deviations of the control model to the actual process, and (4) the convenient application to multivariable control problems. MPC has been considered in the literature for wind turbine load mitigations. A nonlinear MPC (NMPC) method is applied by assuming future wind speed knowledge using a light-detection and ranging (LIDAR) system.\cite{ref:Schlipf2013NMPCLoadsLidar} Simulation results shows promising load reductions without affecting the energy production. Furthermore, a robust MPC (RMPC) implementation is compared to a nominal MPC control structure for the purpose of active tower fore-aft damping.\cite{ref:Evans2015RMPCTowerFA} In numerical simulations, the former outperforms the latter mentioned, as particularly around rated operating conditions, physical actuations constraints form a limiting factor. The benefits of NMPC using a future wind speed prediction are once again emphasized for similar operating conditions.\cite{ref:Tofighi2015NMPCExtremeTransitionConditions}

All of the above described MPC implementations focus on the active mitigation of structural loads. A more passive MPC implementation, providing frequency skipping capabilities, and thereby making an \textit{optimal} trade-off between loads and energy production over the prediction horizon, does not seem to have been backed up by literature in the past. For tall soft-soft tower configurations, the complexity lies in the fact that fatigue loads are minimized by preventing operation at the natural frequency, while it is essential to cross the same frequency for attaining higher rotational speeds and power productions. The conflicting objectives form a burden for describing the objective as a convex optimization problem. Moreover, NMPC for solving nonconvex problems is -- because of its computational complexity -- often considered ineligible for real-time applications.

Imposing spectral constraints might form a possible solution path, by employing the short-time Fourier transform (STFT) on the system output signal in a nonlinear MPC setting.\cite{ref:Hours2015ConstrainedSpectrumControl} A similar methodology\cite{ref:Jain2015MPCModalDamping} uses the selective discrete Fourier transform (SDFT) in an MPC approach to dampen oscillation modes in power system stabilizers (PSS). However, from an implementation and tuning perspective, a frequency domain approach seems to be unintuitive and nontrivial. Therefore, in this paper, another approach is considered. The method involves a model demodulation transformation described for application and control in the field of tapping mode-atomic force microscopy (TM-AFM).\cite{ref:Keyvani2019ModulatedTMAFMModel} The model transformation is applied to the turbine tower model, and transfers frequency-dependent magnitude and phase content to a quasi steady-state contribution. This is accomplished by converting a linear-time invariant (LTI) system description into a linear-parameter varying (LPV) model, scheduled on the excitation frequency. The technique shows similarities with the multiblade coordinate (MBC) transformation\cite{ref:Mulders2019AziOffset}, often used in individual pitch control (IPC) implementations for blade fatigue load reductions.\cite{ref:Bossanyi2003individual,ref:Geyler2007IPC,ref:Mulders2019AziOffset1P2PACC}

An LPV system representation is frequently used for capturing nonlinear dynamics into a system description with a linear input-output mapping.\cite{ref:Hanema2018MPCLPV} An external scheduling variable varies the dynamics of the linear model. Now, consider the combination of an LPV model with MPC. The model-based control method uses a mathematical system description to compute an optimal control signal over the prediction horizon. Unfortunately, for LPV systems, the considered model is subject to changes over time, described by the yet unknown scheduling trajectory. However, when the system is scheduled on state variables and/or input signals, the model is referred to as a quasi\nobreakdash-LPV (qLPV) system. Recently, an efficient MPC scheme for such qLPV systems is proposed by solving subsequent quadratic programs (QPs).\cite{ref:Cisneros2016EfficientNMPC}

This paper subjects a tower model to the earlier introduced demodulation transformation, and augments it with a simplified wind turbine model, such that a qLPV model is obtained. The result is combined with the efficient MPC method, exploiting the beneficial properties of qLPV systems.\cite{ref:Cisneros2016EfficientNMPC} The proposed qLPV\nobreakdash-MPC framework provides a methodology for performing an optimal trade-off between produced energy and tower loads is presented, and thereby presents the following contributions:
\begin{itemize}
	\setlength{\itemsep}{0pt}
	\setlength{\parskip}{0pt}
	\setlength{\parsep}{0pt}
	\item Providing the derivation results of a model demodulation transformation, for moving the magnitude and phase content at the excitation frequency to a quasi steady-state contribution.
	\item Applying the transformation to a second-order tower model, and showcasing its working principles by an illustrative example.
	\item Combining the transformed tower model with a simplified wind turbine model, and linearizing at below-rated operating points, for obtaining a qLPV state-space system description.
	\item Discretizing and converting the qLPV model to its affine form.
	\item Formally deriving the efficient MPC approach for affine qLPV model structures.
	\item Showcasing the proposed approach in closed-loop high-fidelity simulations with different wind profiles, to clearly show its effectiveness and practical applicability.
\end{itemize}

The paper is organized as follows. Section~\ref{sec:AT} describes a methodology for transforming a nominal tower model into a demodulated LPV system description. In Section~\ref{sec:WT}, the obtained system is combined with a simplified wind turbine model, resulting in a qLPV system description after linearization. Next, in Section~\ref{sec:MPC}, the efficient MPC scheme is combined with the qLPV model to make an optimal and user-defined trade-off between tower loads and energy production for the prevailing environmental conditions. In Section~\ref{sec:SIM}, the qLPV\nobreakdash-MPC framework is evaluated with high-fidelity simulations using the NREL~5\nobreakdash-MW reference wind turbine, subject to synthetic and realistic wind profiles. Finally, conclusions are drawn in Section~\ref{sec:CONCLUSIONS}.

\section{Problem formalization and tower model demodulation transformation}\label{sec:AT}
For performing a produced energy versus tower fatigue load trade-off, a wind turbine model needs to be combined with a structural tower model. Section~\ref{sec:AT_TowerModel}, describes the tower side-side dynamics by a second-order mass-damper-spring system. Section~\ref{sec:AT_ProblemFormalization} formalizes the problem statement, and explains why straightforward combination of wind turbine and tower models results in nonconvexity. Therefore, in Section~\ref{sec:AT_Theory}, the nominal tower model is subject to a demodulation transformation to facilitate convexification. The effects and implications of the transformation are analyzed and clarified by an illustrative example in Section~\ref{sec:AT_Example}.
\begin{figure}[t!]
	\centering
	\includegraphics[scale=1.25]{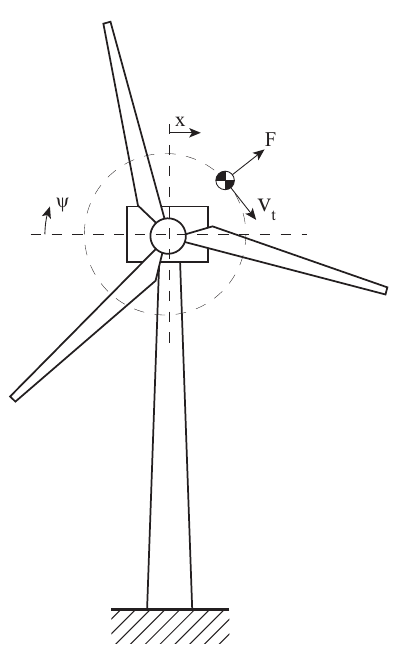}
	\caption{A rotor imbalance excites the turbine support structure because of the centripetal force $F = a_\mathrm{u}\cos{(\psi(t))}$ at the once-per-revolution and rotor-speed dependent ($1$P) frequency. The tangential speed of the imbalance is denoted as $v_\mathrm{t}$, and the side-side tower-top displacement $x$ is given in the hub coordinate system.}
	\label{fig:AT_WTRotorImbalance}
\end{figure}

\subsection{Modeling the tower dynamics as a second-order system}\label{sec:AT_TowerModel}
In practical scenarios, the center of mass of a wind turbine rotor is likely to be unaligned with the rotor center. In effect, as large-scale state-of-the-art wind turbines are operated with a variable-speed control strategy for below-rated conditions, the support structure is excited by a periodic and frequency-varying centripetal force, as illustrated in Figure~\ref{fig:AT_WTRotorImbalance}. The tower dynamics, excited by a rotor-speed dependent once-per-revolution ($1$P) periodic force, are modeled by a second-order mass-damper-spring system
\begin{align}
m\ddot{x}(t) +\zeta\dot{x}(t) + kx(t) &= a_\mathrm{u}\cos{(\psi(t))}\label{eq:AT_NominalModel},
\end{align}
in which ${\left\{ m,\,\zeta,\,k \right\}\in\mathbb{R}^{+}}$ are respectively the constant first mode modal mass, modal damping and modal stiffness, ${\psi(t)\in[0,\,2\pi)}$ is the rotor azimuth angle, ${a_\mathrm{u}\in\mathbb{R}^{+}}$ quantifies the periodic force amplitude, and ${\left\{x,\,\dot{x},\,\ddot{x}\right\}\in\mathbb{R}}$ respectively represent the side-side tower-top displacement, velocity and acceleration in the hub coordinate system, illustrated in Figure~\ref{fig:AT_WTRotorImbalance}. A second-order system is taken to represent the tower first mode using the well known modal-decomposition model reduction technique\cite{ref:Hansen2004StabilityEigenValue, ref:MITModelDecomposition}, and allows for a convenient assessment and derivation of the demodulation transformation in the next section. Application of the transformation to higher-order models is also possible and would result in a similar analysis. Furthermore, the force amplitude $a_\mathrm{u}$ is assumed to be constant for all rotational speeds, however, as will be shown later, this assumption can be relaxed for mildly varying amplitude changes. 

The system in Eq.~\eqref{eq:AT_NominalModel} is split in a set of first-order differential equations by defining $x_1 = \dot{x}(t)$ and $x_2 = x(t)$, respectively representing the tower-top velocity and displacement, such that it is rewritten in the standard state-space $\dot{\boldsymbol{x}} = \mathbf{A_\mathrm{g}}\boldsymbol{x}+\mathbf{B_\mathrm{g}}\boldsymbol{u}$ representation
\begin{align}
\begin{bmatrix}\dot{x}_1\\\dot{x}_2\end{bmatrix}
=\begin{bmatrix}-\mathrm{\zeta/m} & -\mathrm{\omega}_\mathrm{n}^2\\
1 & 0\end{bmatrix}
\begin{bmatrix}x_1\\x_2\end{bmatrix}
+
\begin{bmatrix}\mathrm{a_\mathrm{u}}\\0\end{bmatrix}
\cos{(\psi(t))}\qquad\text{and}\qquad G(s)\stackrel{s}{=}
\left[\begin{array}{c|c}  
	\mathbf{A_\mathrm{g}}&\mathbf{B_\mathrm{g}}\\\hline
	\mathbf{C_\mathrm{g}}&\mathbf{0}
\end{array}\right],\label{eq:AT_NominalModelLin}
\end{align}
in which $\mathbf{A_\mathrm{g}}\in\mathbb{R}^{n_\mathrm{g}\times n_\mathrm{g}}$, $\mathbf{B_\mathrm{g}}\in\mathbb{R}^{n_\mathrm{g}}$, and $\omega_\mathrm{n} = \sqrt{k/m}$ is the structural natural frequency. All states are assumed to be measured, thus $\mathbf{C_\mathrm{g}} = I_\mathrm{n_g}$. Using the operator $\stackrel{s}{=}$ to equate the state-space system description to the transfer function with Laplace variable $s$, is a notation taken from Skogestad \textit{et. al}.\cite{ref:Skogestad2007multivariable} The notation means that the transfer function $G(s)$ has a state-space realization given by the quadruple $\left(\mathbf{A_\mathrm{g}},\,\mathbf{B_\mathrm{g}},\,\mathbf{C_\mathrm{g}},\,\mathbf{0}\right)$. An explicit definition of the transfer function is omitted in this work, since the notation is only used as a convenient way of parameterizing and referring to the state-space system.

\subsection{Problem formalization}\label{sec:AT_ProblemFormalization}
\begin{figure}[t!]
	\centering
	\includegraphics[scale=1.0]{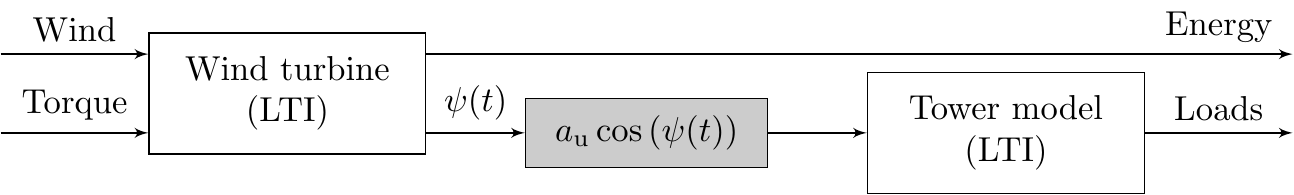}\vspace{0.2cm}\\ $\mathrel{\scalebox{1}[2]{$\downarrow$}}$ \\\vspace{0.2cm}
	\includegraphics[scale=1.0]{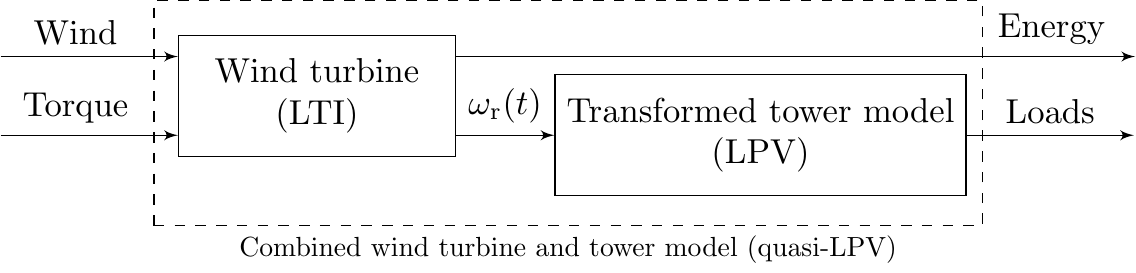}
	\caption{In the \textit{upper diagram}, a wind turbine model is driven by wind disturbance and torque control inputs, and has energy production and azimuth position as outputs. The latter mentioned output is taken by a cosine function, and serves as an input to the tower model, resulting in a fatigue load signal. The presence of the trigonometric function forms a barrier for describing the energy-load trade-off as a convex optimization problem.  Therefore, in the \textit{lower diagram}, the trigonometric function is combined with the tower model by a model demodulation transformation, resulting in an LPV system description. Joining the wind turbine model with the transformed tower model results in a quasi\nobreakdash-LPV system description, in which the rotor speed state serves as the scheduling variable.}
	\label{fig:AT_ProblemFormalization}
\end{figure}

This section formalizes the problem considered in this paper. The aim is to provide a trade-off between energy production efficiency and tower fatigue load reductions, by preventing rotor speed operation near the tower natural frequency. The considered nominal framework is graphically presented in~Figure~\ref{fig:AT_ProblemFormalization}. The wind turbine model has a wind disturbance and a generator torque control input, the latter of which is subject to optimization. A cosine function acts on the azimuth position output from the wind turbine model, which results in a periodic input to the tower model.

The load and energy outputs of the respective tower and wind turbine models, together with the torque input signal are included in the following cost function to optimize the energy-load trade-off
\begin{align}
\underset{Torque}{\mathrm{arg\,min}}\quad & -\lambda_\mathrm{1}(Energy) + \lambda_\mathrm{2}(Loads) + \lambda_\mathrm{3}(Torque),\label{eq:AT_OptimizationProblem}
\end{align}
in which $\lambda_i~\text{and}~i = \{1,2,3\}$ are positive weighting constants determining the objective trade-offs. The above given relation presents the optimization objective in an informal fashion for illustration purposes; later sections define the problem in a mathematical correct way. The load signal is a periodic and rotor-speed dependent measure for tower fatigue loading, caused by the presence of the trigonometric function. This forms a burden for describing the objective as a convex optimization problem. 

The solution path employed in this paper is to subject the combined nonlinear trigonometric function and LTI tower model by a demodulation transformation. The transformed tower model results in an LPV system description. The subsequent aggregation with a wind turbine model results in a quasi\nobreakdash-LPV model, as the internal rotor speed state serves as the scheduling variable. The next section provides theory for derivation of the demodulation transformation, whereas later sections elaborate on an efficient MPC method exploiting the beneficial properties of qLPV model structures. The process of obtaining a demodulated version of a periodically excited LTI model, is referred to as the demodulation transformation in the remainder of this paper.

\subsection{Theory on the tower model demodulation transformation with periodic excitation towards an LPV representation}\label{sec:AT_Theory}
In signal analysis, \textit{modulation} is the process of imposing an information-bearing signal onto a second signal, often referred to as the carrier, carrier signal or carrier wave.\cite{ref:Oppenheim2013SignalsSystems} Subsequently, the procedure of recovering the signal of interest from the modulated signal is called \textit{demodulation}. The modulation-demodulation scheme is in communication systems mostly performed using dedicated hardware. A second method for performing the operation is the derivation of a mathematical framework for obtaining a demodulated model, which can be applied when the dominant dynamics of the nominal process are known. The latter mentioned approach is employed in this work, as it will appear useful for analysis, controller design, and forward propagation in MPC.

The aim of the demodulation transformation is to obtain a linear (but parameter varying) system description, which provides the frequency dependent dynamical behavior as a steady-state signal. The demodulated signal is used in a later section to form a convex quadratic optimization problem in an MPC setting for computing the optimal control signal over the prediction horizon. The demodulation transformation is applied to the assumed tower model, introduced in Section~\ref{sec:AT_TowerModel}.

\begin{figure}[t!]
	\centering
	\begin{tabular}{m{0.45\linewidth} c m{0.45\linewidth}}
		\includegraphics[height=1.1cm]{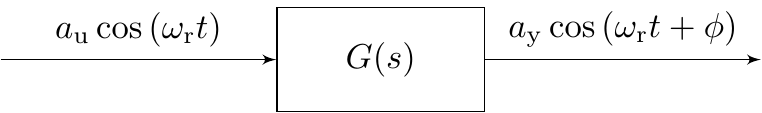} &\qquad
		&\includegraphics[height=2.1cm]{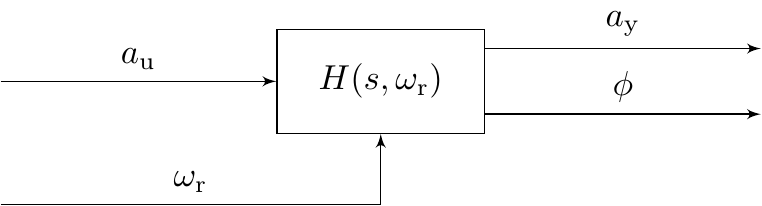}
	\end{tabular}
	\caption{\textbf{\textit{Left:}} the nominal tower model is periodically excited at a certain frequency and amplitude. For the linear case, the response is scaled and phase-shifted with respect to the driving input signal. \textbf{\textit{Right:}} after the demodulation transformation, the input amplitude is a direct input to the system, whereas its frequency changes the system dynamics. The resulting outputs give the response amplitude and phase shift as a quasi\nobreakdash-steady state signal.}
	\label{fig:AT_NominalAmplitudeTransformation}
\end{figure}

The model demodulation transformation is inspired by the work of\cite{ref:Keyvani2019ModulatedTMAFMModel}, and only the main results are given in this section. The derivation is performed and validated with symbolic manipulation software for algebraic expressions\cite{ref:MATLAB2018bSymbolic}, and the code is made publicly available.\cite{ref:Mulders2019DataScriptsZenodo} The transformation relies on the assumption that the changes in system response amplitude $a_\mathrm{y}(\tau)$ and phase change $\phi(\tau)$ are much slower than that of the driving excitation frequency $\omega_\mathrm{r}$. For this reason, the slower time scale is indicated by $\tau$ as a substitute for the normal time scale $t$. Variables that are a function of the slow-varying time scale are assumed to be constant over a single period $T_\mathrm{r} = 2\pi/\omega_\mathrm{r}$ of the excitation:
\begin{align}
\int^{T_\mathrm{r}}_{0} f(\tau)g(t)dt &= f(\tau)\int^{T_\mathrm{r}}_{0}g(t)dt\label{eq:AT_IntegralProperty}.
\end{align}
Driving a linear system with a periodic input, results in a periodic response with the same frequency, however, with a certain phase shift and magnitude relative to that of the input, which is characterized by:
\begin{align}
x_i(t) &= a_i(\tau)\cos{(\omega_\mathrm{r} t + \phi(\tau))},
\end{align}
in which $\phi\in\mathbb{R}$ is the phase shift, $a_i\in\mathbb{R}^+$ is the amplitude, and the subscript $i\in\mathbb{Z}^+$ is a counter variable. By taking into account the introduced time-scales, and using Euler's formula $e^{\mathrm{j}\phi} = \cos(\phi) + \mathrm{j}\sin{(\phi)}$, the state variables are rewritten as
\begin{align}
x_i(t) &= \Re{\left\{a_i(\tau)e^{\mathrm{j}(\omega_\mathrm{r} t+\phi(\tau))}\right\}},
\end{align}
with $\mathrm{j}=\sqrt{-1}$ being the imaginary unit, and $i = \{1,2\}$, where $i=1$ relates to velocity and $i=2$ to displacement. The symbol $\Re\{\cdot\}$ indicates the real part of a given expression, whereas $\Im{\{\cdot\}}$ is used to represent the imaginary part. The slow varying term ${X_i\in\mathbb{C}}$ is now written as a product with the fast harmonic function, with a fixed phase and amplitude:
\begin{align}
x_i(t) & = \Re{\left\{a_i(\tau)e^{\mathrm{j}\phi(\tau)}e^{\mathrm{j}\omega_\mathrm{r} t}\right\}} = \Re{\left\{X_i(\tau)e^{\mathrm{j}\omega_\mathrm{r} t}\right\}},\label{eq:T_StateReal}
\end{align}
and taking the first time derivative gives
\begin{align}
\dot{x}_i(t) = \Re{\left\{ \left(\dot{X}_i(\tau)+ \mathrm{j}\omega_\mathrm{r} X_i(\tau) \right)e^{\mathrm{j}\omega_\mathrm{r} t}\right\}}.\label{eq:T_StateTimeDerReal}
\end{align}
By substitution of Eqs.~\eqref{eq:T_StateReal}~and~\eqref{eq:T_StateTimeDerReal} in the nominal state space representation of Eq.~\eqref{eq:AT_NominalModelLin}, the following expressions are obtained
\begin{align}
&\Re\left\{\left( \dot{X}_1(\tau) + j\omega_\mathrm{r} X_1(\tau) + (\zeta/m)X_1(\tau) + \omega_\mathrm{n}^2X_2(\tau) \right)e^{j\omega_\mathrm{r} t}- a_\mathrm{u}e^{j\omega_\mathrm{r} t}\right\} = 0,\label{eq:T_SubsRealInStateSpace1}\\
&\Re{\left\{\left( \dot{X}_2(\tau) + j\omega_\mathrm{r} X_2(\tau) - X_1(\tau) \right)e^{j\omega_\mathrm{r} t}\right\}} = 0.\label{eq:T_SubsRealInStateSpace2}
\end{align}
Furthermore, the following property of orthogonality is used, where
\begin{align}
\int_{0}^{T_\mathrm{r}} \Re{\left\{ Ce^{j\theta} \right\}}e^{j\theta}d\theta = 0,
\end{align}
if and only if $\{C\in\mathbb{C}\} = 0$. Thus, Eqs.~\eqref{eq:T_SubsRealInStateSpace1}~and~\eqref{eq:T_SubsRealInStateSpace2} are multiplied with $e^{\mathrm{j}\omega_\mathrm{r} t}$ as follows:
\begin{align}
&\int_{0}^{T_\mathrm{r}} \Re\left\{\left( \dot{X}_1(\tau) + j\omega_\mathrm{r} X_1(\tau) + (\zeta/m)X_1(\tau) + \omega_\mathrm{n}^2X_2(\tau) \right)e^{j\omega_\mathrm{r} t}- a_\mathrm{u}e^{j\omega_\mathrm{r} t}\right\}e^{\mathrm{j}\omega_\mathrm{r} t}dt = 0,\\
&\int_{0}^{T_\mathrm{r}} \Re{\left\{\left( \dot{X}_2(\tau) + j\omega_\mathrm{r} X_2(\tau) - X_1(\tau) \right)e^{j\omega_\mathrm{r} t}\right\}}e^{\mathrm{j}\omega_\mathrm{r} t}dt = 0.\label{eq:T_IntQ12}
\end{align}
Term-by-term integration of the integrals in Eq.~\eqref{eq:T_IntQ12} using the mathematical property in Eq.~\eqref{eq:AT_IntegralProperty}, gives the following result
\begin{align}
\begin{bmatrix} \dot{X}_1 \\ \dot{X}_2 \end{bmatrix} &=
\begin{bmatrix}
\mathrm{j}\omega_\mathrm{r} X_1 - (\mathrm{\zeta}/\mathrm{m})X_1-\omega_\mathrm{n}^2X_2 + a_\mathrm{u} \\
\mathrm{j} \omega_\mathrm{r} X_2 + X_1
\end{bmatrix}.
\end{align}
Now, by defining a new state sequence $\boldsymbol{q} = [q_1,\,q_2,\,q_3,\,q_4]^\mathrm{T} = [\Re\{X_1\},\,\Im\{X_1\},\,\Re\{X_2\},\,\Im\{X_2\}]^\mathrm{T}$, the system is rewritten as ${\boldsymbol{\dot{q}}(\omega_\mathrm{r}) = \mathbf{A}_\mathrm{h}(\omega_\mathrm{r})\boldsymbol{q} + \mathbf{B}_\mathrm{h}}$, such that the following expression is obtained:
\begin{align}
\begin{bmatrix}
\dot{q}_1\\
\dot{q}_2\\
\dot{q}_3\\
\dot{q}_4
\end{bmatrix}=
\begin{bmatrix}
-\mathrm{\zeta}/\mathrm{m} & \omega_\mathrm{r} & -\omega_\mathrm{n}^2 & 0\\
-\omega_\mathrm{r} & -\mathrm{\zeta}/\mathrm{m} & 0 & -\omega_\mathrm{n}^2\\
1 & 0 & 0 & \omega_\mathrm{r}\\
0 & 1 & -\omega_\mathrm{r} & 0
\end{bmatrix}
\begin{bmatrix}
{q}_1\\
{q}_2\\
{q}_3\\
{q}_4
\end{bmatrix}+
\begin{bmatrix}
a_\mathrm{u}\\
0\\
0\\
0
\end{bmatrix}\qquad\text{and}\qquad H(s, \omega_\mathrm{r})\stackrel{s}{=}
\left[\begin{array}{c|c}  
	\mathbf{A_\mathrm{h}}&\mathbf{B_\mathrm{h}}\\\hline
	\mathbf{C_\mathrm{h}}&\mathbf{0}
\end{array}\right],
\label{eq:StateSpace_Transformed}
\end{align}
in which $\mathbf{A_\mathrm{h}}\in\mathbb{R}^{n_\mathrm{h}\times n_\mathrm{h}}$, $\mathbf{B_\mathrm{h}}\in\mathbb{R}^{n_\mathrm{h}}$ and $n_\mathrm{h} =  2n_\mathrm{g}$. Again, all states are measured, thus $\mathbf{C_\mathrm{h}} = I_\mathrm{n_h}$. The system $H(s,\omega_\mathrm{r})$ has a state-space realization given by the quadruple $\left(\mathbf{A_\mathrm{h}},\,\mathbf{B_\mathrm{h}},\,\mathbf{C_\mathrm{h}},\,\mathbf{0}\right)$.\cite{ref:Skogestad2007multivariable} The instantaneous amplitude and phase of the dynamic system response at frequency $\omega_\mathrm{r}$ are given by
\begin{align}
a_\mathrm{y}(\tau) &= \sqrt{q_3^2+q_4^2},\label{eq:AT_Ay}\\
\phi(\tau) &= \tan^{-1}{(q_4/q_3)}.
\end{align}
It is also possible to write the result of the derivation using a summation of Kronecker products
\begin{align}
	H(s,\omega_\mathrm{r}) &\stackrel{s}{=} \boldsymbol{\dot{q}} = 
	\left(\mathbf{A}_\mathrm{g}\otimes
	I_\mathrm{n_g}
	+
	I_\mathrm{n_g}\otimes
	\begin{bmatrix}
		0 & \omega_\mathrm{r} \\ -\omega_\mathrm{r} & 0
	\end{bmatrix}\right)\boldsymbol{q} + \left(\mathbf{B_\mathrm{g}}\otimes\begin{bmatrix}
		1\\0
	\end{bmatrix}\right).
\end{align}
The nominal and transformed model representations $G(s)$ and $H(s,\omega_\mathrm{r})$ are interchangeable: Figure~\ref{fig:AT_NominalAmplitudeTransformation} graphically summarizes the transformation of the nominal periodically excited second-order tower model (Eq.~\eqref{eq:AT_NominalModelLin}) into an LPV model structure (Eq.~\eqref{eq:StateSpace_Transformed}). The amplitude $a_\mathrm{u}$ of the periodic input is in the demodulated model a direct input to the system. The outputs are the amplitude $a_\mathrm{y}$ and phase shift $\phi$ with respect to the input frequency. Note that the frequency $\omega_\mathrm{r}$ is in the transformed case a scheduling variable to the LPV system, changing the system dynamics. The following section demonstrates and further explains the effects of the presented transformation by an illustrative analysis.

\subsection{Illustrating the effects of the transformation}\label{sec:AT_Example}
This section adds context to the rather abstract derivation of the demodulation transformation in Section~\ref{sec:AT_Theory}. Therefore, the first part of this section presents a frequency domain analysis of the transformation properties. Then, in an illustrative time domain simulation case, a frequency sweep is applied to the nominal and transformed models. The analysis and exemplary simulation clarify the characteristics and applicability of the transformation for the considered objective in this paper.
\begin{figure}[b!]
	\centering
	\includegraphics[scale=1.0]{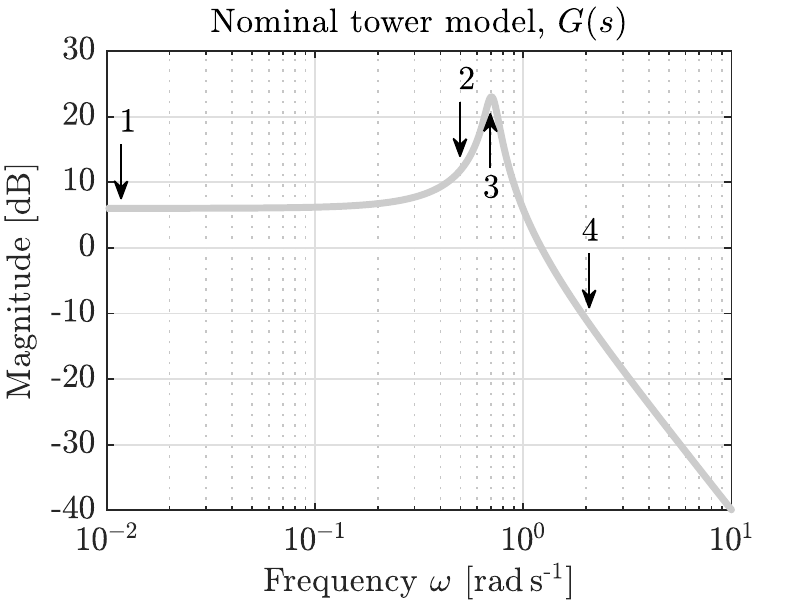}
	\includegraphics[scale=1.0]{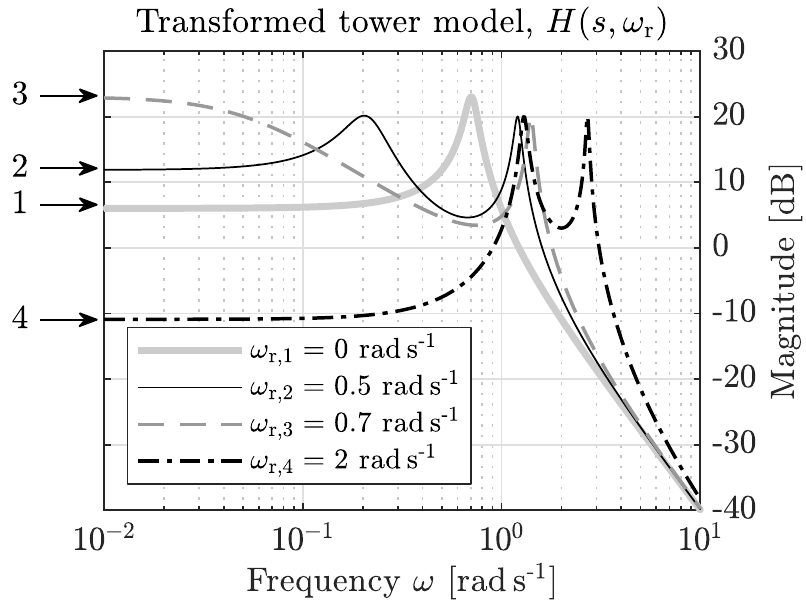}
	\caption{\textbf{\textit{Left:}} Frequency response of the nominal tower model $G(\mathrm{j}\omega)$. A clear tower resonance peak is observed at $\omega_\mathrm{n} \approx 0.71$~rad\,s\textsuperscript{-1}, and a $-40$~dB/decade roll-off at higher frequencies. A set of 4 comparison points $\omega_{\mathrm{r},i}\in\boldsymbol{\Omega_\mathrm{r}} = \left\{0,\,0.5,\,0.7,\,2.0\right\}$~rad\,s\textsuperscript{-1} is chosen for evaluation of the nominal and demodulated model. \textbf{\textit{Right:}} Frequency responses of the transformed model $H(\mathrm{j}\omega, \omega_{\mathrm{r},i})$ for the set of comparison points. The magnitude content at the indicated frequencies in the left plot is transferred to a steady-state contribution in the transformed case. When the input signal $a_\mathrm{u}$ to the transformed model is considered constant or slowly varying, the additional resonances at higher frequencies do not contribute to the output.}
	\label{fig:AT_FRFNormalAT}
\end{figure}

The nominal and transformed tower models from Eqs.~\eqref{eq:AT_NominalModelLin}~and~\eqref{eq:AT_Ay} are parameterized by the following quantities: A modal mass of $m = 1000$~kg, a modal damping coefficient of $\zeta = 100$~kg\,s\textsuperscript{-1}, and a modal spring constant of $k = 500$~kg\,s\textsuperscript{-2}. Resulting from the somewhat arbitrarily selected parameters, the first tower mode is located at $\omega_\mathrm{n} \approx 0.71$~rad\,s\textsuperscript{-1}, with a clearly present resonance peak at the same frequency. Section~\ref{sec:SIM} modifies the NREL~5\nobreakdash-MW reference turbine tower to move its side-side fundamental frequency to the same location. Figure~\ref{fig:AT_FRFNormalAT} shows Bode magnitude plots of the nominal plant and its demodulated counterpart in the frequency range ${\boldsymbol{\Omega}=\left\{ \omega\,|\,\omega\subset\mathbb{R},~10^{-2}\leq\omega\leq10^{1}~rad\,s\textsuperscript{-1}\right\}}$. To obtain the amplitude output $a_\mathrm{y}$ of the demodulated model, the Euclidean norm of the frequency responses of $q_3$ and $q_4$ at each frequency point is taken. 

Figure~\ref{fig:AT_FRFNormalAT} showcases the frequency domain effects of the transformation. The frequency responses are evaluated for 4 rotor speeds $\omega_{\mathrm{r},i}\in\boldsymbol{\Omega_\mathrm{r}}$, $i = \left\{1,\,2,\,3,\,4\right\}$, defined by the set $\boldsymbol{\Omega_\mathrm{r}} = \left\{0,\,0.5,\,0.7,\,2.0\right\}~\text{rad\,s\textsuperscript{-1}}$. The rotor speed elements parameterize the transformed model $H(s,\omega_\mathrm{r})$. The plots in Figure~\ref{fig:AT_FRFNormalAT} show arrows, indicating that frequency dependent magnitude information (left) is transferred to a steady-state contribution (right). Note that for $\omega_{\mathrm{r},1} = 0$~rad\,s\textsuperscript{-1} the transformed model reduces to the nominal case. Moreover, the right plot shows that the nominal resonance peak at $\omega_\mathrm{n}$ is for each frequency response split into two peaks with a $3$~dB magnitude reduction. In effect, when the input amplitude $a_\mathrm{u}$ of the transformed model is constant or varied slowly, the magnitude at specific nominal model frequency points is mapped to a DC contribution in the transformed case; rapid variations will result in contributions from the resonances at higher frequencies. However, in this paper, additional measures to reduce these effects, such as low-pass or notch filters, are dispensable, because $a_\mathrm{u}$ is assumed to be constant.

\begin{figure}[t!]
	\centering
	\includegraphics[scale=1.0]{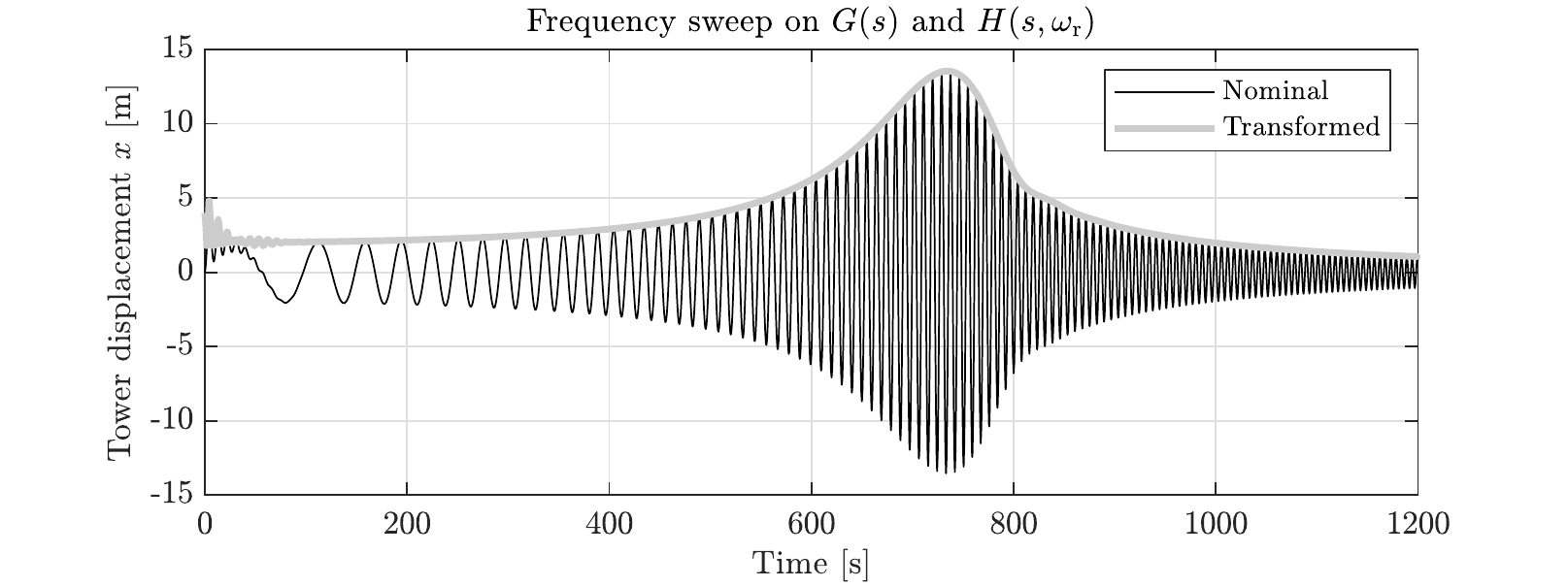}
	\caption{Frequency sweep applied to the nominal and transformed model, from $\omega_\mathrm{r} =0$ to $1.2$~rad\,s\textsuperscript{-1} with a constant acceleration in $1200$~s. The transformed model shows a very close amplitude tracking of the nominal model magnitude response.}
	\label{fig:AT_FrequencySweep}
\end{figure}

Figure~\ref{fig:AT_FrequencySweep} shows the time-domain characteristics of the transformation. For this, a frequency sweep is applied to the nominal and transformed models. For the total simulation time of $1200$~seconds, the signal has a linearly increasing frequency, with a constant increase rate of ${\dot{\omega}_\mathrm{r} = 10^{-3}}$~rad\,s\textsuperscript{-2}, starting from ${\omega_\mathrm{r} = 0}$ to $1.2$~rad\,s\textsuperscript{-1}. This frequency range is chosen as modern large-scale variable-speed wind turbines are controlled in this operating region. The transformed model shows a very close amplitude tracking of the nominal model dynamics. The earlier imposed assumption on the change in amplitude and phase by a slow time scale $\tau$, does not seem to limit the proposed method for applicability to the considered wind turbine control objective.

\section{Wind turbine model augmentation and linearization}\label{sec:WT}
This section considers the derivation of a simple (linear) NREL~5\nobreakdash-MW model, for augmentation to the demodulated tower model such that a quasi\nobreakdash-LPV model is obtained. Section~\ref{sec:WT_Model} provides the simple first-order wind turbine model. Next, in Section~\ref{sec:WT_Linear}, the model is symbolically linearized and augmented with the transformed tower model in a qLPV representation. Section~\ref{sec:WT_CompletingLin} provides linearization parameters over the complete below-rated operating region based on the properties of the NREL~5\nobreakdash-MW reference wind turbine.\cite{ref:Jonkman2009DefinitionNREL5MW} Finally, Section~\ref{sec:WT_Example} validates the first-order and affine linear models to simulation results of their nonlinear equivalent.

\subsection{Simplified wind turbine system description}\label{sec:WT_Model}
Because the dynamics of the transformed tower model $H(s, \omega_{\mathrm{r}})$ are scheduled by the input excitation frequency, which is in this case the ($1$P) rotor speed, it is a logical step to augment a wind turbine model adding this state to the overall system description. A system of which the scheduling variable is part of the state vector is known as a qLPV system description. The considered first-order wind turbine model is
\begin{align}
J_\mathrm{r}\dot{\omega}_\mathrm{r} &= \tau_\mathrm{a} - \underbrace{N(\tau_\mathrm{g} + \Delta\tau_\mathrm{g})}_{\tau_\mathrm{s}},\label{eq:WT_FistOrderModel}
\end{align}
in which $J_\mathrm{r}\in\mathbb{R}^+$ is the total rotor inertia consisting out of the hub and 3 times the blade inertia, $\left\{N\geq 1\right\} \subset\mathbb{R}^+$ is the gearbox ratio, and $\tau_\mathrm{a}$ is the aerodynamic rotor torque defined as
\begin{align}
	\tau_\mathrm{a} &= \frac{1}{2}\rho_\mathrm{a}\pi R^3 U^2 C_\mathrm{\tau}(\lambda, \beta),\label{eq:WT_AeroDynamicTorque}
\end{align}
in which $\rho_\mathrm{a}\in\mathbb{R}^+$ is the air density, $R\in\mathbb{R}^+$ the rotor radius, $U\in\mathbb{R}^+$ the rotor effective wind speed, and $C_\mathrm{\tau}\in\mathbb{R}$ the torque coefficient as a function of the blade pitch angle $\beta$ and the dimensionless tip-speed ratio $\lambda = \omega_\mathrm{r}R/U$. The system torque $\tau_\mathrm{s}\in\mathbb{R}^+$ is a summation of the generator torque $\tau_\mathrm{g}\in\mathbb{R}^+$ resulting from a standard \textit{K-omega-squared} torque control strategy\cite{ref:Bossanyi200DesignFeedback}, and $\Delta\tau_\mathrm{g}\in\mathbb{R}$ is an additional torque contribution resulting from the MPC framework described later in this paper. The \textit{K-omega-squared} torque control law is taken as an integral part of the model, and is defined as
\begin{align}
	\tau_\mathrm{g} = K\omega_\mathrm{r}^2/N,\label{eq:WT_Komega2}
\end{align}
in which $K\in\mathbb{R}^+$ is the optimal mode gain 
\begin{align}
K = \frac{\pi\rho_\mathrm{a}R^{5}C_\mathrm{p}(\lambda,\beta)}{2\lambda^3},
\end{align}
calculated for the low-speed shaft (LSS) side.

\subsection{Linearizing the augmented turbine and tower model}\label{sec:WT_Linear}
This section augments the wind turbine model from Section~\ref{sec:WT_Model} to the demodulated tower model $H(s,\omega_\mathrm{r})$, such that the following system is obtained:
\begin{align}
\begin{bmatrix}
\dot{q}_1\\
\dot{q}_2\\
\dot{q}_3\\
\dot{q}_4\\
\dot{\omega}_\mathrm{r}
\end{bmatrix}&=
\begin{bmatrix}
-\mathrm{\zeta}/\mathrm{m} & \omega_\mathrm{r} & -\omega_\mathrm{n}^2 & 0 & 0\\
-\omega_\mathrm{r} & -\mathrm{\zeta}/\mathrm{m} & 0 & -\omega_\mathrm{n}^2 & 0\\
1 & 0 & 0 & \omega_\mathrm{r} & 0\\
0 & 1 & -\omega_\mathrm{r} & 0 & 0\\
0 & 0 & 0 & 0 & 0
\end{bmatrix}
\begin{bmatrix}
{q}_1\\
{q}_2\\
{q}_3\\
{q}_4\\
\omega_\mathrm{r}
\end{bmatrix}+
\begin{bmatrix}
a_\mathrm{u}\\
0\\
0\\
0\\
(\tau_\mathrm{a} - N\left(\tau_\mathrm{g}+\Delta\tau_\mathrm{g}\right))/J_\mathrm{r}
\end{bmatrix},\label{eq:WT_AugmentedNonLinSysIn}\\
a_\mathrm{y} &= \sqrt{(q_3^2 + q_4^2)}\label{eq:WT_AugmentedNonLinSysOut}.
\end{align}
The above-given system description contains the nonlinear aerodynamic and generator torque input defined previously by Eqs.~\eqref{eq:WT_AeroDynamicTorque}~and~\eqref{eq:WT_Komega2}. Furthermore, the output $a_\mathrm{y}$ is a nonlinear combination of state vector elements. The system is subject to linearization, where the desired linear state, input and output vectors are defined as
\begin{align}
\boldsymbol{\hat{q}}(t) &= \left[ \hat{q}_1,\,\hat{q}_2,\,\hat{q}_3,\,\hat{q}_4,\,\hat{\omega}_\mathrm{r} \right]^{\rm T},\\
\boldsymbol{\hat{u}}(t) &= \left[ \hat{U},\,\Delta\hat{\tau}_\mathrm{g} \right]^{\rm T},\\
\boldsymbol{\hat{y}}(t) &= \hat{A}_\mathrm{y},
\end{align}
and the $\hat{(\cdot)}$-notation indicates the deviation with respect to the considered linearization point. Now, the system is linearized by taking the partial derivatives of Eqs.~\eqref{eq:WT_AugmentedNonLinSysIn}~and~\eqref{eq:WT_AugmentedNonLinSysOut} with respect to the state and inputs vectors, such that a linear state-space system is obtained
\begin{align}
\boldsymbol{\dot{\hat{q}}}(t)&=\mathbf{A}(\boldsymbol{p})\boldsymbol{\hat{q}}(t)+
\mathbf{B}(\boldsymbol{p})\boldsymbol{\hat{u}}(t)\label{eq:WT_AugmentedLinSys}\\
\boldsymbol{\hat{y}}(t) &=
\mathbf{C}(\boldsymbol{p})\boldsymbol{\hat{q}}(t)\nonumber,
\end{align}
in which the state, input and output matrices are defined as
\begin{align}
\mathbf{{A}}(\boldsymbol{p}) &= \begin{bmatrix}
-\mathrm{\zeta}/\mathrm{m} & \bar{\omega}_\mathrm{r} & -\omega_\mathrm{n}^2 & 0 & \bar{q}_2\\
-\bar{\omega}_\mathrm{r} & -\mathrm{\zeta}/\mathrm{m} & 0 & -\omega_\mathrm{n}^2 & -\bar{q}_1\\
1 & 0 & 0 & \bar{\omega}_\mathrm{r} & \bar{q}_4\\
0 & 1 & -\bar{\omega}_\mathrm{r} & 0 & -\bar{q}_3\\
0 & 0 & 0 & 0 & (\bar{k}_\mathrm{\omega_\mathrm{r}}-N\bar{k}_{\tau_\mathrm{g}})/J_\mathrm{r}
\end{bmatrix},\quad
\mathbf{B}(\boldsymbol{p}) = \begin{bmatrix}
0 & 0\\
0 & 0\\
0 & 0\\
0 & 0\\
\bar{k}_\mathrm{U}/J & -N/J
\end{bmatrix},\quad
\mathbf{C}(\boldsymbol{p}) = \frac{1}{2}\begin{bmatrix}
0 \\ 0 \\ \bar{q}_3(\bar{q}_3^2 + \bar{q}_4^2)^{-1/2} \\ \bar{q}_4(\bar{q}_3^2 + \bar{q}_4^2)^{-1/2} \\ 0
\end{bmatrix}^{\mathrm{T}}.
\end{align}
The aerodynamic rotor torque is linearized with respect to the rotor speed and wind speed
\begin{align}
\hat{\tau}_\mathrm{a} &= \frac{\partial\tau_\mathrm{a}}{\partial\omega_\mathrm{r}}\hat{\omega}_\mathrm{r} + \frac{\partial\tau_\mathrm{a}}{\partial U}\hat{U} = \bar{k}_\mathrm{\omega_\mathrm{r}}({\omega_\mathrm{r}},~{\beta},~{U})\hat{\omega}_\mathrm{r} + \bar{k}_\mathrm{U}(\omega_\mathrm{r},\beta,U)\hat{U},
\end{align}
with
\begin{align}
\bar{k}_\mathrm{\omega_\mathrm{r}}(\omega_\mathrm{r},\beta,U) &= \left.c_\mathrm{r}RU\frac{\partial C_\mathrm{\tau}(\lambda,\beta)}{\partial\lambda}\right|_{\omega_\mathrm{r} = \bar{\omega_\mathrm{r}},\,\beta = \bar{\beta},\,U = \bar{U}},\label{eq:WT_LinKW}\\
\bar{k}_\mathrm{U}(\omega_\mathrm{r},\beta,U) &= \left.2c_\mathrm{r}C_\mathrm{\tau}(\lambda,\beta)U-c_\mathrm{r}\omega_\mathrm{r}R\frac{\partial C_\mathrm{\tau}(\lambda,\beta)}{\partial\lambda}\right|_{\omega_\mathrm{r} = \bar{\omega}_\mathrm{r},\,\beta = \bar{\beta},\,U = \bar{U}},\label{eq:WT_LinKU}
\end{align}
and $c_\mathrm{r} = 0.5\rho\pi R^3$ is a constant factor. Finally, the \textit{K-omega-squared} torque controller is linearized as
\begin{align}
\hat{\tau}_\mathrm{g}(\omega_\mathrm{r}) &= \left.\frac{\partial\tau_\mathrm{g}}{\partial\omega_\mathrm{r}}\hat{\omega}_\mathrm{r} = \bar{k}_{\tau_\mathrm{g}}(\omega_\mathrm{r})\hat{\omega}_\mathrm{r} = 2K{\omega_\mathrm{r}}/N\right|_{\omega_\mathrm{r} = \bar{\omega}_\mathrm{r}}\hat{\omega}_\mathrm{r}.\label{eq:WT_LinTG}
\end{align}
The $\bar{\left(\cdot\right)}$-notation indicates the steady-state values of the corresponding operating points. The advantage of this approach is that for each operating point, corresponding steady-state values are substituted in the state-space matrices. This is done by a function $\boldsymbol{p} = f(\omega_\mathrm{r}(t)):\mathbb{R}\rightarrow \mathbb{R}^{n_\mathrm{p}}$, which schedules the system $\mathbf{A}(\boldsymbol{p}):\mathbb{R}^{n_\mathrm{p}}\rightarrow\mathbb{R}^{n\times n}$, input $\mathbf{B}(\boldsymbol{p}):\mathbb{R}^{n_\mathrm{p}}\rightarrow\mathbb{R}^{n\times m}$ and output $\mathbf{C}(\boldsymbol{p}):\mathbb{R}^{n_\mathrm{p}}\rightarrow\mathbb{R}^{q\times n}$ matrices. This leads to the description of nonlinear dynamics by a set of linear models, varying the system according to the operating point parameterized by $\boldsymbol{p}\in\mathcal{P}$. For the qLPV case, the scheduling variable is part of the state, which makes the system self-scheduling at each time step. In this paper, a finite number of linearizations is considered for operating conditions along the optimal power coefficient $C_\mathrm{p,max}(\lambda^*) = C_\mathrm{\tau}(\lambda^*)\lambda^*$ corresponding to the set $\boldsymbol{\mathcal{U}}$ of below-rated wind speeds. 

The current form of the linear model in Eq.~\eqref{eq:WT_AugmentedLinSys} only describes deviations from the current operating point. To approach the actual states and outputs of the nonlinear model with a qLPV model structure, offsets for the state, input and output should be incorporated in the system description. The process of incorporating these operating point offsets, converting the LPV model to its affine form, is described in Appendix~\ref{sec:APP_Affine}. The same appendix also describes the employed fourth order Runge-Kutta state-space discretization method. When in the remainder of this paper is referred to the qLPV model, the system in its affine form is intended.

\subsection{Completing the linearization for the NREL~5-MW reference turbine}\label{sec:WT_CompletingLin}
This section provides the data for linearization of the NREL~5\nobreakdash-MW turbine, and performs a validation of the resulting affine qLPV system to the nonlinear turbine model in high-fidelity simulation code. All linearization parameters are summarized in Table~\eqref{tab:WT_TableSimParameters}.
\begin{table}[t!]
	\caption{Parameters for linearization and simulation of the qLPV model in the below-rated operating region.}
	\label{tab:WT_TableSimParameters}
	\begin{center}
		\begin{tabular}{lccc}
			\hline
			\textbf{Description} & \textbf{Symbol} & \textbf{Value} & \textbf{Unit}\\
			\hline\hline
			Blade inertia & $J_\mathrm{b}$ & $11.776\cdot 10^6$ & kg\,m\textsuperscript{2}\\
			Hub inertia & $J_\mathrm{h}$ & $115\,926$ &  kg\,m\textsuperscript{2}\\
			Total rotor inertia & $J_\mathrm{h}$ & $35.444\cdot 10^6$ &  kg\,m\textsuperscript{2}\\
			Torque coefficient fit (1/2) & $\theta_{1,2}$ & $\left.14.5924,\,42.7653\right.$ & -\\
			Torque coefficient fit (2/2) & $\theta_{3,4}$ & $\left.2.4604,\,0.0036\right.$ & - \\
			Gearbox ratio & $N$ & $97$ & -\\ 
			Air density & $\rho_\mathrm{a}$ & $1.225$ & kg\,m\textsuperscript{-3}\\
			Fine pitch angle & $\beta_0$ & $1.9\cdot10^{-3}$ & rad\\
			Rotor radius & $R$ & $63$ & m\\
			Optimal mode gain (LSS) & $K$ & $2.1286\cdot10^{6}$ & Nm\,(rad\,s\textsuperscript{-1})\textsuperscript{-2}\\
			Optimal tip-speed ratio & $\lambda^*$ & $7.7$ & -\\ 
			Input excitation amplitude & $a_\mathrm{u}$ & $1$ & -\\			
			Tower mass & $m$ & $1000$ & kg\\
			Tower damping & $\zeta$ & $100$ & kg\,s\textsuperscript{-1}\\ 
			Tower stiffness & $k$ & $500$ & kg\,s\textsuperscript{-2}\\ 
			Tower natural frequency & $\omega_\mathrm{n}$ & $0.7071$ & rad\,s\textsuperscript{-1}\\ 
			\hline
		\end{tabular}
	\end{center}
\end{table}

\begin{figure}[t!]
	\centering
	\includegraphics[scale=1.0]{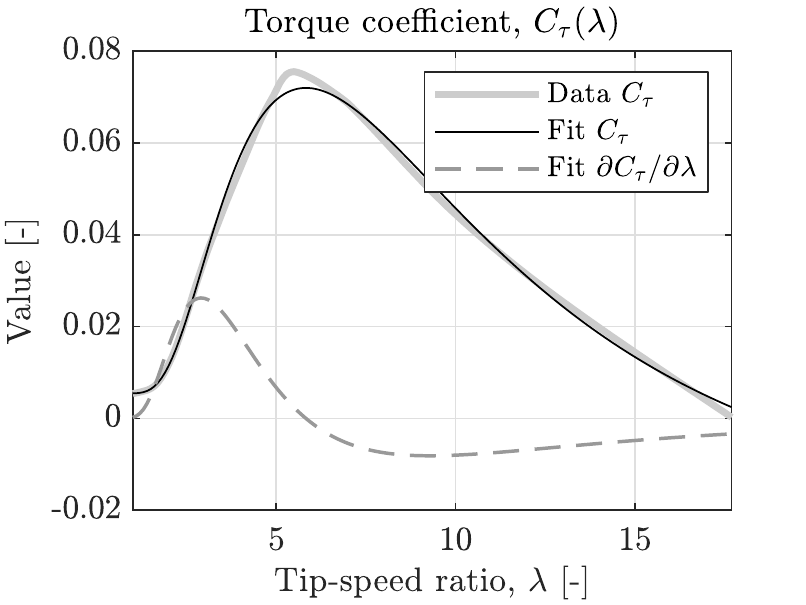}
	\includegraphics[scale=1.0]{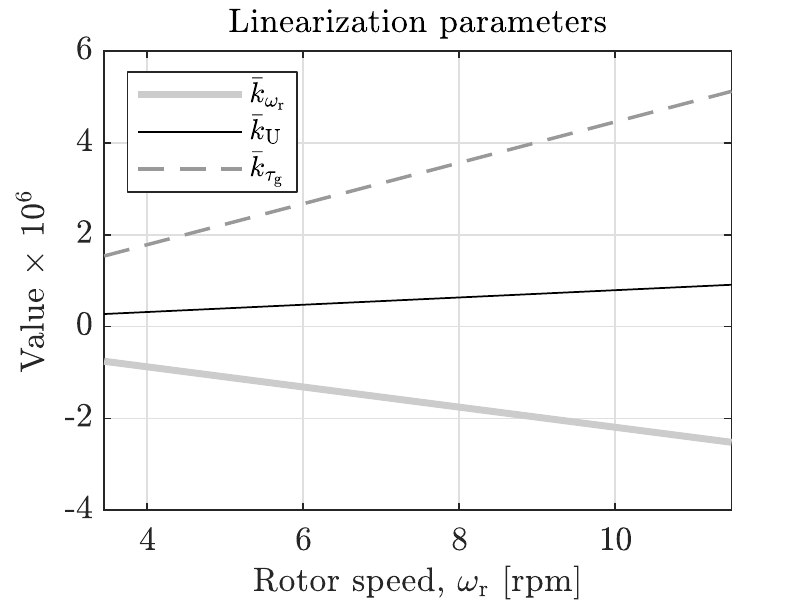}
	\caption{\textbf{\textit{Left:}} torque coefficient curve of the NREL~5\nobreakdash-MW reference wind turbine as a function of the dimensionless tip-speed ratio. The fit according to the model structure proposed by Eq.~\eqref{eq:WT_CTauModelStructure} shows a close fit to the data points. The fit allows the derivation and evaluation of the partial gradient. \textbf{\textit{Right:}} the linearization parameters defining the LPV model at each scheduling instant.}
	\label{fig:WT_CQDataFit}
\end{figure}

First, an analytical fit is made to the NREL~5\nobreakdash-MW torque coefficient data as a function of the tip-speed ratio. This is needed as $\bar{k}_\mathrm{\omega_r}$ and $\bar{k}_\mathrm{U}$ are a function of the operational rotor and wind speed. The torque coefficient data is obtained using a graphical extension\cite{ref:FASTv8GUI} to NREL's high-fidelity wind turbine simulation software FAST v8.16\cite{ref:FASTv816}, which includes blade element momentum (BEM) code\cite{ref:Burton2001WindEnergyHandbook} for obtaining rotor characteristic data. As the framework being derived in this paper focuses on the below-rated region, and conventional wind turbine controllers keep the pitch angle fixed at fine-pitch angle $\beta_0$ during partial load\cite{ref:Pao2011ControlOfWTs}, the dependency of the torque coefficient on $\beta$ is omitted. An often used parameterizable torque coefficient function is defined by
\begin{align}
C_\mathrm{\tau}(\lambda) &= {\rm e}^{-\theta_{1}/\lambda}(\theta_{2}/\lambda-\theta_{3})/\lambda+\theta_{4} \label{eq:WT_CTauModelStructure},
\end{align}
which is fitted by optimizing the values $\theta_i$ using a nonlinear least-squares routine, minimizing the sum-of-squares between the fit and the data-points. Figure~\ref{fig:WT_CQDataFit} shows the torque coefficient trajectory as a function of the tip-speed ratio for $\beta = \beta_0$, and the fit to this data. Also, an evaluation of the analytically computed partial gradient with respect to the tip-speed ratio is given. Furthermore, the same figure shows the linearization parameters $\bar{k}_\mathrm{\omega_r}$, $\bar{k}_\mathrm{U}$ and $\bar{k}_\mathrm{\tau_g}$. The evaluation is performed for all below-rated rotor speed conditions along the maximum power coefficient $C_\mathrm{p,max}$ at an optimal tip-speed ratio of $\lambda^* = 7.7$. The trajectories show smooth and linear behavior for all operating points.
\begin{figure}[t!]
	\centering
	\includegraphics[scale=1.0]{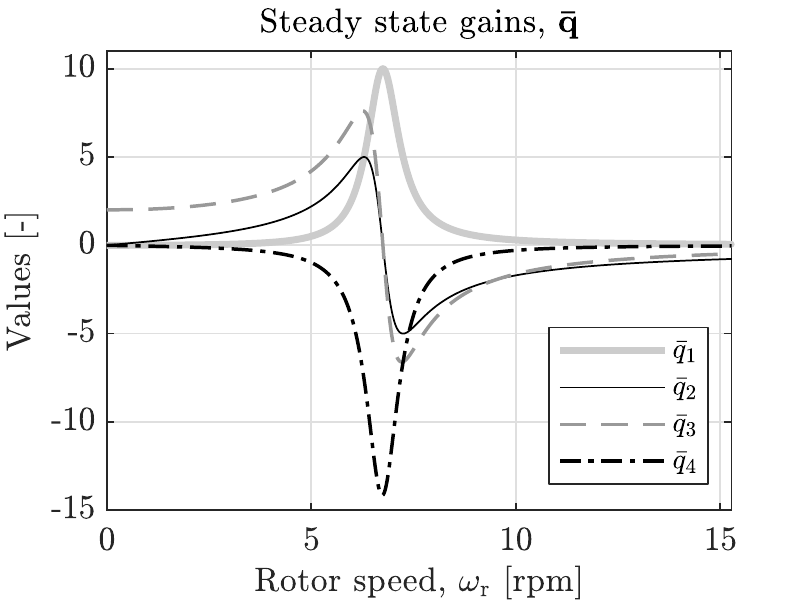}
	\caption{State-state gains $\bar{q}_1$ to $\bar{q}_4$ as a function of the rotor speed scheduling variable. Around the natural tower frequency, the gains show a higher sensitivity to the x-coordinate, raising the need for an LPV model set on a fine scheduling grid.}
	\label{fig:WT_ssGainsq1234}
\end{figure}

In Figure~\ref{fig:WT_ssGainsq1234}, the steady-state values for $\bar{q}_1$, $\bar{q}_2$, $\bar{q}_3$, and $\bar{q}_4$ are given as a function of rotor speed for the optimal power coefficient operating conditions. Compared to the previously presented linearization parameters, these trajectories show a more volatile behavior: At the tower natural frequency, two of the trajectories change signs, while the other two reach their extrema. This, as will be shown later, results in some erratic behavior when the qLPV model self-schedules itself around the natural frequency. Therefore, a fine grid of linear models should be taken in the LPV scheduling space, to minimize artifacts and to properly describe the nonlinear dynamics.

\subsection{The qLPV model subject to a turbulent wind}\label{sec:WT_Example}
\begin{figure}[t!]
	\centering
	\includegraphics[scale=1.0]{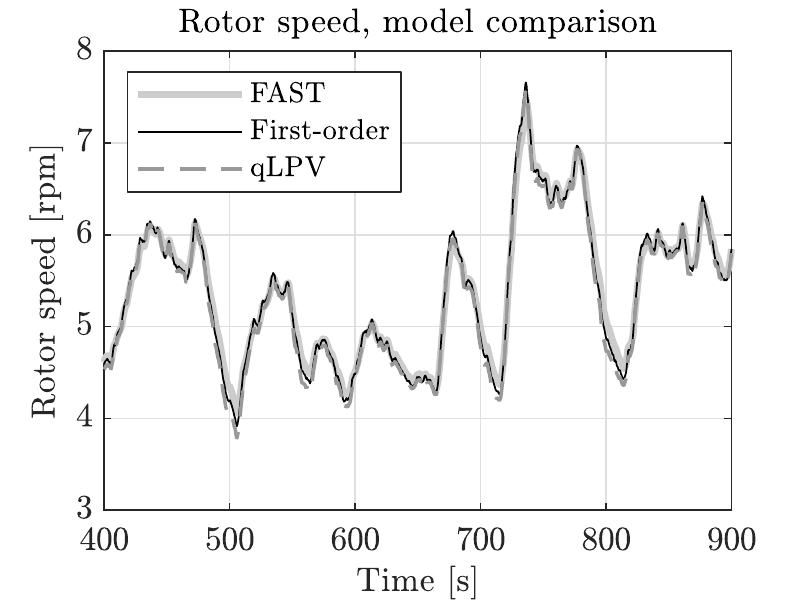}
	\includegraphics[scale=1.0]{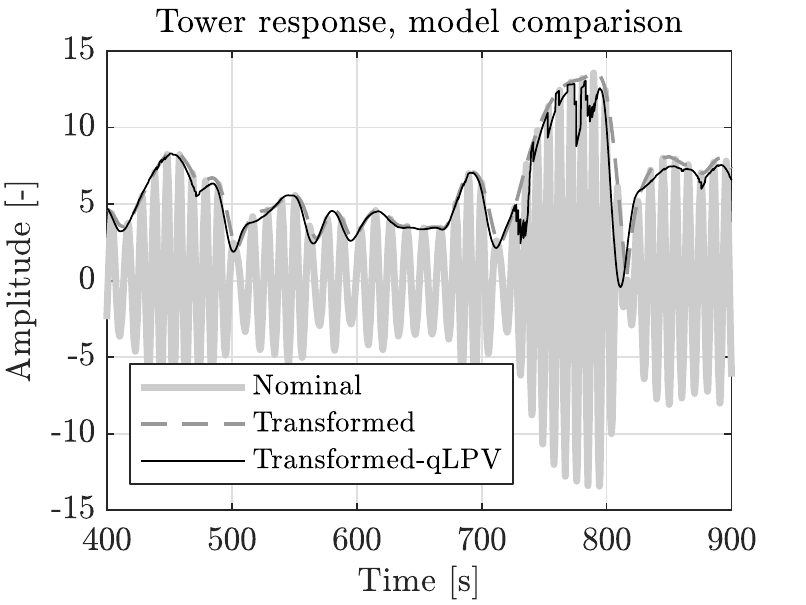}
	\caption{Simulation results showing the rotor speed and tower-top displacement amplitude for different models driven by a turbulent wind disturbance input signal. The FAST model excites the nominal tower model and serves as a baseline simulation case. The first-order wind turbine model is combined with the nonlinearized transformed tower model. The qLPV system is a single scheduled system description for the rotor and tower dynamics, and shows -- apart from minor artifacts around the tower natural frequency -- a close resemblance to its nonlinear companion.}
	\label{fig:WT_Simulation_RotSpeedAmplitude}
\end{figure}

The main advantage of a qLPV model, is that the scheduling parameter is part of the state vector. In this way, the scheduling signal is not exogenous, and the model is consequently self-scheduling according to its state evolution. To verify the validity of the derived affine qLPV model, a turbulent wind signal with a mean wind speed of $\bar{U} = 6.5\,$m\,s\textsuperscript{-1} is applied to:
\begin{enumerate}
	\setlength{\itemsep}{0pt}
	\setlength{\parskip}{0pt}
	\setlength{\parsep}{0pt}
	\item A nonlinear NREL~5\nobreakdash-MW aerodynamic model simulated in the high-fidelity \textbf{FAST} code. A second-order, first mode tower model $G(s)$ is excited by a unity amplitude cosine as a function of the azimuth position.
	\item The \textbf{first-order} linearized NREL~5\nobreakdash-MW wind turbine model with $C_\mathrm{\tau}(\lambda)$ look-up table, driving the \textbf{transformed} tower model $H(s,\omega_\mathrm{r})$ by the rotor speed output.
	\item The \textbf{qLPV} model, incorporating the linear wind turbine rotor and \textbf{transformed} tower dynamics, self-scheduled by its rotor speed state.
\end{enumerate}
The simulation results, based on the parameters in Table~\eqref{tab:WT_TableSimParameters}, are presented in Figure~\ref{fig:WT_Simulation_RotSpeedAmplitude}. The left plot shows the rotor speed simulation, which demonstrates that the first-order and qLPV models accurately follow the FAST output. Subsequently, the right plot compares the tower-top side-side displacement responses, as a result of the rotor imbalance excitation. As concluded earlier by the frequency sweep in Figure~\ref{fig:AT_FrequencySweep}, the nominal and transformed tower models show a good match. The additional qLPV response in Figure~\ref{fig:WT_Simulation_RotSpeedAmplitude} shows a similar trajectory as the transformed model, apart from some minor artifacts between $700-800$~s, when the rotor speeds approaches the tower natural frequency. These anomalies are a result of the steady-state gains $\bar{q}_{1-4}$ being more sensitive in the region of $\omega_\mathrm{n}$ (Figure~\ref{fig:WT_ssGainsq1234}), and the switching between the linear models by the scheduling parameter. However, as the response serves as a load indication, and the exact value is of less importance, the qLPV method is concluded being suitable for its intended purpose in the qLPV\nobreakdash-MPC framework, described in the next section.

\section{Quasi-LPV model predictive control}\label{sec:MPC}
Nonlinear MPC is -- because of its computational complexity -- often considered as an unsuitable control method for application in fast real-time systems. Therefore, in this paper, an approach towards an efficient method for nonlinear MPC is employed, exploiting the inherent self-scheduling property of a qLPV system. This section describes the qLPV\nobreakdash-MPC framework, with the aim to provide a convex QP, defining a trade-off between maximizing power production efficiency, and minimizing tower natural frequency excitation. Practically, this means that the rotor deviates from the maximum power extraction trajectory when it approaches the rotor speed coinciding with the structural resonance frequency.

An economic MPC approach is used to directly optimize for the economic performance of the process.\cite{ref:Rawlings2012FundamentalsEcoMPC} For the considered case, a predefined quadratic performance criterion specifies the trade-off between energy production maximization and resonance excitation minimization, and the optimizer finds the optimal corresponding control signal in the prediction horizon. However, as for each time step the scheduling sequence over the prediction horizon is unknown, the nonlinear MPC control problem is solved by an iterative method.\cite{ref:Cisneros2016EfficientNMPC} The method solves subsequent QPs minimizing the predefined cost, and uses the resulting predicted scheduling sequence as a warm-start for the next iteration. Each iteration uses a single QP solve. A norm on the consecutive output differences is used to determine whether the algorithm has converged. 

By manipulation of the affine system representation defined by Eqs.~\eqref{eq:WT_AffineDiscLPVStateEq}~and~\eqref{eq:WT_AffineDiscLPVOutputEq}, an expression is derived for forward propagation of the qLPV model output, only requiring the initial state at time instant $k$ and the scheduling sequence over the prediction horizon:
\begin{align}
\boldsymbol{Y}_{k+1} &= \mathbf{H}(\mathbf{P}_{k})\left(x_{k} - \breve{x}(p_\mathrm{k})\right) + \mathbf{S}(\mathbf{P}_{k}) \Delta\boldsymbol{U}_{k}(\mathbf{P}_{k}) + \left(\boldsymbol{\breve{Y}}_{k+1}(\mathbf{P}_{k}) +\mathbf{L}(\mathbf{P}_{k}) \Delta\boldsymbol{\breve{X}}_{k}(\mathbf{P}_{k}) + \mathbf{D}(\mathbf{P}_{k}) \Delta\boldsymbol{U}_{k+1}(\mathbf{P}_{k})\right),\label{eq:WT_MPC_Equation}
\end{align}
in which the matrices $\left\{\boldsymbol{H},\,\boldsymbol{S},\,\Delta\boldsymbol{U}_{k},\,\boldsymbol{\breve{Y}}_{k+1},\,\mathbf{L},\,\Delta\boldsymbol{\breve{X}}_{k},\,\mathbf{D},\,\Delta\boldsymbol{U}_{k+1}\right\}$ are defined in Appendix~\ref{sec:APP_ForwardPropagation}, and ${\mathbf{P}_\mathrm{k} = \left[\boldsymbol{p}_{k},\,\boldsymbol{p}_{k+1}~\cdots~\boldsymbol{p}_{k+\mathrm{N_p}}\right]\in\mathbb{R}^{n_\mathrm{p}\times N_\mathrm{p}}}$ is the collection of scheduling variables at each time instant over the prediction horizon ${N_\mathrm{p}\in\mathbb{Z}^+}$. The $\breve{(\cdot)}$-notation indicates steady-state offsets from the current operating point for the the states, in- and outputs (Appendix~\ref{sec:APP_Affine}). The opportunity for defining a control horizon ${N_\mathrm{c}\in\mathbb{Z}}$ is disregarded in this paper, and is chosen to equal $N_\mathrm{p}$. For sake of completeness, the above given propagation expression includes a direct feedthrough matrix $\mathbf{D}$, although it is not used for the considered problem. 

At time instant $k = 0$, only the initial state is assumed to be known, and the scheduling parameters are chosen constant over the prediction horizon, such that:
\begin{align}
\boldsymbol{P_\mathrm{0}} &= \boldsymbol{1}_\mathrm{N_p}\otimes\boldsymbol{p}_\mathrm{0}\label{eq:MPC_InitSchedP0},
\end{align}
in which ${\boldsymbol{1}_\mathrm{N_p}\in\mathbb{R}^{N_\mathrm{p}}}$ is a one-dimensional vector of ones. By assuming the initialization vector, the convex QP is solved with ${\Delta\boldsymbol{\Theta}_{\mathrm{g},k+1} = \left[\Delta\tau_{\mathrm{g},k+1}\cdots\Delta\tau_{\mathrm{g},k+N_\mathrm{p}}\right]\in\mathbb{R}^{N_\mathrm{p}}}$ as the decision variable vector, minimizing the cost
\begin{align}
\underset{\Delta\boldsymbol{\Theta}_{\mathrm{g},k+1}}{\mathrm{arg\,min}} \quad & \boldsymbol{Y}_{k+1}^T\mathbf{Q}\boldsymbol{Y}_{k+1} + \Delta\boldsymbol{\Theta}_{\mathrm{g},k+1}^T\mathbf{R}\Delta\boldsymbol{\Theta}_{\mathrm{g},k+1}\label{eq:MPC_ArgMin}\\
\text{subject to}\quad & \text{Dynamical system in Eq.~}\eqref{eq:WT_MPC_Equation},\nonumber
\end{align}
in which ${\mathbf{Q} = \texttt{diag}(Q,\,Q~\cdots~Q)\in\mathbb{R}^{\mathrm{N_p}\times\mathrm{N_p}}}$ and ${\mathbf{R} = \texttt{diag}(R,\,R~\cdots~R)\in\mathbb{R}^{\mathrm{N_p}\times\mathrm{N_p}}}$ are, respectively, weight matrices acting on the predicted tower-top displacement amplitude and deviation from the optimal torque control signal. The latter term of the cost requires the assumption of optimal power production efficiency using the \textit{K-omega-squared} torque control strategy. Now, compare the above given minimization objective with the one introduced in the problem formalization by Eq.~\eqref{eq:AT_OptimizationProblem}. The first term of Eq.~\eqref{eq:MPC_ArgMin} aims on fatigue load minimization, whereas the latter term is a combination of energy production maximization and penalization on the control input. Formulating the objective in this way, results in a convenient trade-off between power production and load reductions by varying the weight ratio of $Q$ and $R$.
\begin{algorithm}[t!]
\caption{-~Pseudocode for iteratively finding the scheduling vector $\mathbf{P}_\mathrm{k}$ in the first time step, and warm-starting for subsequent time instants.}\label{alg:MPC_qLPV-MPC_Iterations}
\begin{algorithmic}
	\State $k\leftarrow0$,\quad$j\leftarrow1$,\quad$j_\mathrm{n}\leftarrow5$	
	\State Define $Q$, $R$, $N_\mathrm{p}$
	
	\State Initialize matrices $\mathbf{Q}$, $\mathbf{R}$
	\State Initialize state $\mathbf{X}$, output $\mathbf{Y}$, and scheduling $\boldsymbol{P}$ matrices as empty $\mathbf{0}$-matrices
	\State $\mathbf{P}^{j}_{k} \leftarrow  1_\mathrm{N_p}\otimes f(\boldsymbol{x}_{0})$,\quad$\mathbf{X}(:,k)= \boldsymbol{x}_{0}$
	
	\For {time instant $k$}
	
	\While {$j \leq j_\mathrm{n}$}
	\State Construct matrices $\mathbf{H}(\mathbf{P}^{j}_{k})$, $\mathbf{S}(\mathbf{P}^{j}_{k})$, $\mathbf{L}(\mathbf{P}^{j}_{k})$, $\Delta\boldsymbol{U}(\mathbf{P}^{j}_{k})$, $\boldsymbol{\breve{Y}}_{k+1}(\mathbf{P}^{j}_{k})$,  $\Delta\boldsymbol{\breve{X}}_{k+1}(\mathbf{P}^{j}_{k})$
	\State Solve for $\Delta\boldsymbol{\Theta}_\mathrm{g}$ as in Eq.~\eqref{eq:MPC_ArgMin} with $\mathbf{X}(:,k)$ as initial state
	\State Simulate the qLPV model with $\Delta\boldsymbol{\Theta}_\mathrm{g}$ for $N_\mathrm{p}$ samples to find the state evolution $\boldsymbol{\mathcal{X}}^{j}$
	\State Define $\mathbf{P}^{j+1}_{k} = f(\boldsymbol{\mathcal{X}}^{j})$ 
	\State $j \leftarrow j + 1$
	\EndWhile
	\State $j \leftarrow 1$,\quad$j_\mathrm{n}\leftarrow 1$
	\State Take the first sample of $\Delta\boldsymbol{\Theta}_\mathrm{g}$ to apply in high-fidelity code and simulate for $t_\mathrm{s,FAST}$
	\State Save resulting state and output data: $\mathbf{X}(:,k+1)\leftarrow \boldsymbol{x}_{k}$,\quad$\mathbf{Y}(:,k)\leftarrow \boldsymbol{y}_{k}$
	\State Define $\mathbf{P}^{j}_{k+1} = f(\boldsymbol{\mathcal{X}}^{\mathrm{end}})$ as a warm start for the next time instant
	\State $k \leftarrow k + 1$
	\EndFor
\end{algorithmic}
\end{algorithm}
\begin{figure}[t!]
	\centering
	\includegraphics[scale=1.0]{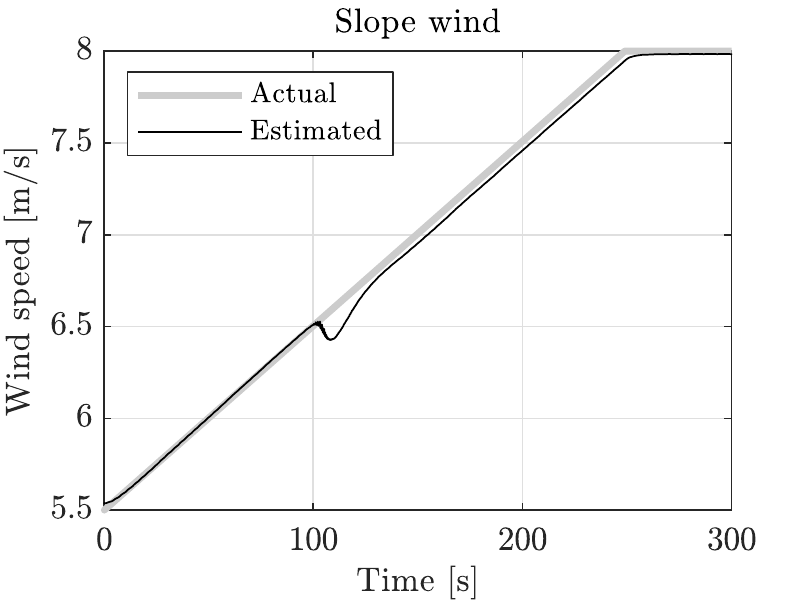}
	\includegraphics[scale=1.0]{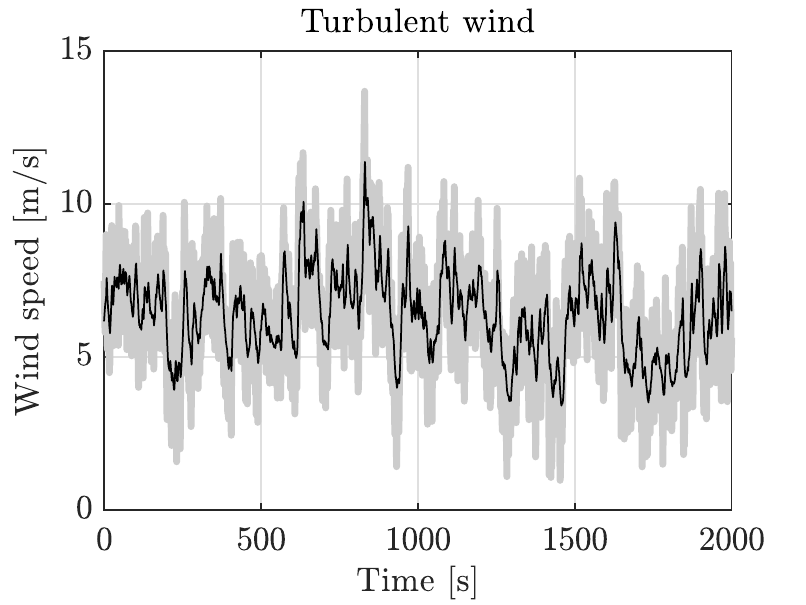}
	\caption{A linearly increasing slope and turbulent wind profile employed for the two simulation cases. The rotor effective wind speed is estimated by a wind speed estimator. A discrepancy of the estimated sloped wind speed is observed after $100$~s, which is a result of sudden changes in applied generator torque and measured rotor speed.}
	\label{fig:SIM_Wind}
\end{figure}

After the first solve with the initial scheduling sequence of Eq.~\eqref{eq:MPC_InitSchedP0}, the inherent qLPV property is exploited by using the predicted evolution of the state to form a warm-start initialization of $\mathbf{P}_{k}^{j+1}$ in the next iteration. This iterative process is repeated until $\left|\left|\boldsymbol{Y}_{k}^{j+1} - \boldsymbol{Y}_{k}^{j}\right|\right|_{2} < \epsilon$, or for a maximum number of iterations $j_\mathrm{n}$, with $\epsilon$ being a predefined error threshold. The algorithm is summarized using pseudocode in Algorithm~\ref{alg:MPC_qLPV-MPC_Iterations}.
\begin{figure}[t!]
	\centering
	\includegraphics[scale=1.0]{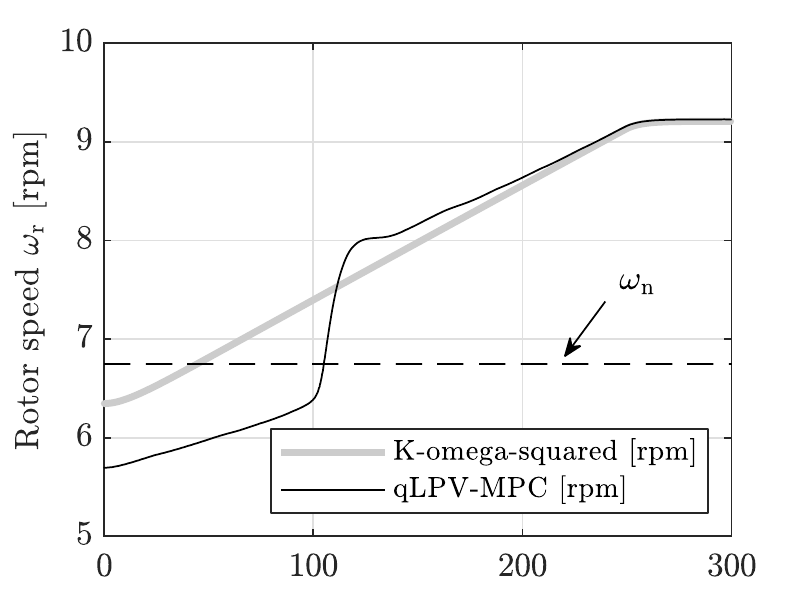}
	\includegraphics[scale=1.0]{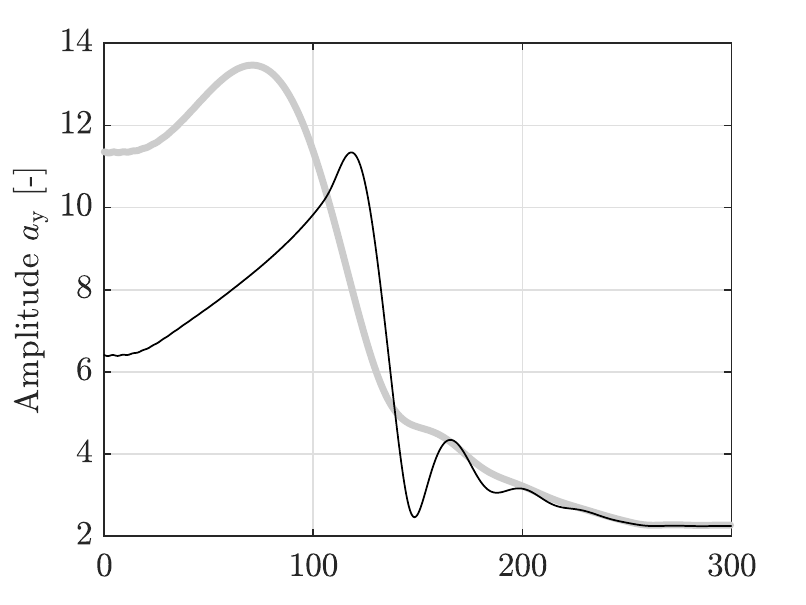}
	\includegraphics[scale=1.0]{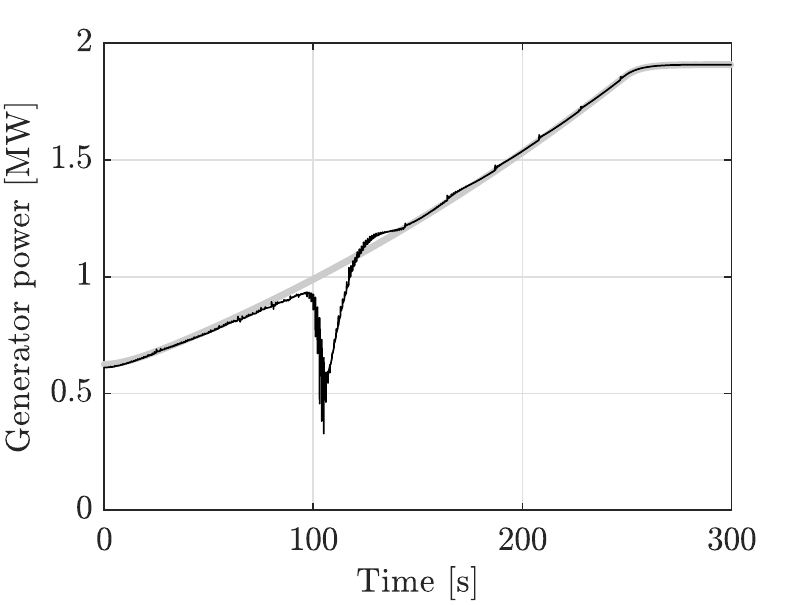}
	\includegraphics[scale=1.0]{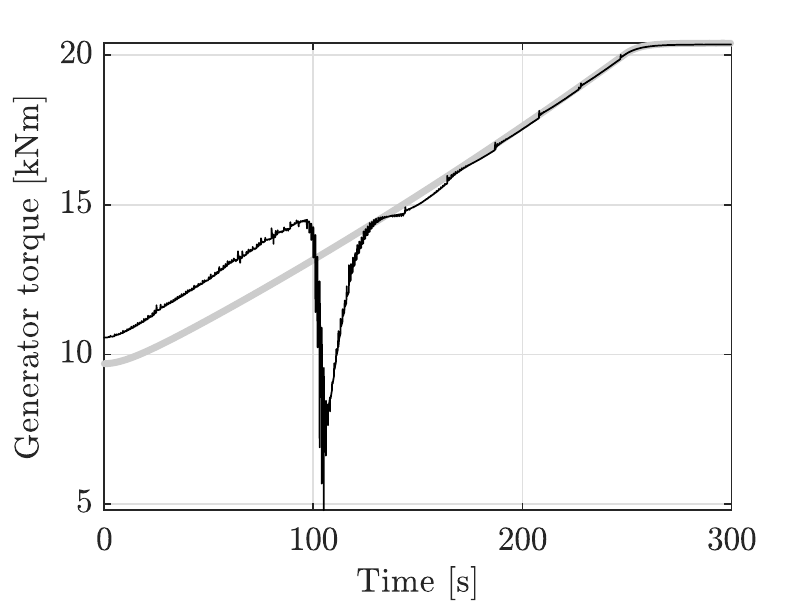}
	\caption{Simulation \textbf{Case 1} shows a comparison with conventional torque control subject to a linearly increasing wind speed. The proposed algorithm prevents the rotor speed from prolonged operation at the tower's natural frequency by imposing an additional generator torque demand. Then, when the wind speed is sufficient for operation at a higher rotor speed, the additional generator torque is rapidly reduced to facilitate a swift crossing of the critical frequency. The strategy is beneficial for reducing periodic tower loading at a specific frequency, at the expense of generated power.}
	\label{fig:SIM_Slope}
\end{figure}

An evaluation has shown that after convergence during initialization, warm-starting the scheduling sequence for the subsequent time-steps shows excellent results. That is, performing multiple iterations for time instants $k>0$ shows no significant performance enhancements for the considered problem. Therefore, the described process is only performed in the initial time step $k = 0$. The need for only a single QP in each step, makes the approach for solving the nonlinear MPC problem computationally efficient and tractable for real-world implementations.

\section{High-fidelity simulation setup and results}\label{sec:SIM}
This section implements the proposed qLPV\nobreakdash-MPC framework in conjunction with the NREL~5\nobreakdash-MW reference wind turbine model in the high-fidelity FAST code. The software implementation is made publicly available.\cite{ref:Mulders2019DataScriptsZenodo} As the side-side natural frequency of the NREL~5\nobreakdash-MW turbine is located outside the rotor speed operating region, the tower properties are modified. The tower wall thickness is scaled down by a factor $7.5$ to mimic the characteristics of a tall, more flexible, soft-soft tower configuration. As a result of the reduced thickness, an effective turbine side-side resonance frequency of approximately $0.71$~rad\,s\textsuperscript{-1} is attained, equal to $\omega_\mathrm{n}$ defined in Section~\ref{sec:AT_Example}. Also, two of the three blades are configured to have an overall mass increase and decrease of $2$~\% with respect to the reference blade. This mass imbalance induces a rotor eccentricity, exacerbating the excitation of the turbine side-side mode.

Furthermore, the simulation environment incorporates the demodulated second-order tower model from Eq.~\eqref{eq:StateSpace_Transformed}. The transformed tower model is scheduled by the simulated rotor speed, and the resulting integrator states are, together with the rotor speed, used in each time-step to form the initial state. The initial state is, as shown in Algorithm~\ref{alg:MPC_qLPV-MPC_Iterations}, at each time instant used for forward propagation of the qLPV model by Eq.~\eqref{eq:WT_MPC_Equation}.

The aim is now to showcase the framework capabilities of successfully preventing prolonged rotor speed operation near the tower resonance frequency. This is done by defining two separate simulation cases:
\begin{itemize}
	\setlength{\itemsep}{0pt}
	\setlength{\parskip}{0pt}
	\setlength{\parsep}{0pt}
	\item \textbf{Case 1:} Initializing the wind turbine for operating conditions corresponding to a wind speed of $U = 5.5$~m\,s\textsuperscript{-1}, followed by a linearly increasing slope to a maximum wind speed of $U = 8.0$~m\,s\textsuperscript{-1} in approximately $250$~s.
	\item \textbf{Case 2:} Operating the wind turbine in turbulent wind conditions with a mean wind speed $\bar{U} = 6.5$~m\,s\textsuperscript{-1} for $2000$~s.
\end{itemize}
For both cases, the behavior of the qLPV\nobreakdash-MPC implementation is compared with standard \textit{K-omega-squared} torque control.

The employed wind signals are presented in Figure~\ref{fig:SIM_Wind}. Because the wind speed cannot assumed to be measurable in real-world scenarios, an effective immersion and invariance (I\&I) rotor effective wind speed estimator is used\cite{ref:Ortega2013WindSpeedEstimator,ref:Soltani2013estimation}, which is also plotted in the same figure. Because the future wind speed is unknown at time instant $k$, the wind speed evolution is chosen to be constant and equal to the current estimated value over the prediction horizon. Also, the smoothened course of the estimated signal aids the qLPV\nobreakdash-MPC algorithm to prevent from overreacting to rapid variations. As the wind speed estimator takes the applied generator torque and measured rotor speed as inputs, and a rapid rotor speed and generator torque change occurs after $100$~s, a discrepancy is seen at this time instant. Nonetheless, the estimator shows a quick recovery in consequent time steps. 

Distinct sampling intervals are used for the simulation environment and MPC update actions. Simulation of the NREL~5\nobreakdash-MW reference turbine in FAST requires a sampling time of $t_\mathrm{s,FAST} = 0.01$~s to prevent numerical issues. The MPC sampling time is set to $t_\mathrm{s,MPC} = 1.0$~s. Note that this rather low sampling interval is possible because the demodulation transformation moves the load signal to a quasi\nobreakdash-steady state contribution. As a result of this transformation, the algorithm's goal is to find the optimal operating trajectory, and not to actively mitigate a specific frequency. The low sampling interval is especially convenient for real-world applications, as this allows solving the QP less frequently, reducing the need for powerful control hardware.

The FAST simulation environment, implemented in MATLAB Simulink\cite{ref:MATLAB2018bSimulink}, simulates for $t_\mathrm{s,MPC}$, after which is simulation is paused, and essential information is extracted. The simulation data is provided to the MPC algorithm in MATLAB using CVX: A package for specifying and solving convex programs.\cite{ref:Boyd2014CVX,ref:Boyd2008NonsmoothConvex} After solving the optimization problem, the first input sample of the decision variable vector is updated in the simulation environment. The simulation is resumed and the input is held constant for the next $t_\mathrm{s,MPC}$ seconds, after which it is paused again.
\begin{figure}[t!]
	\centering
	\includegraphics[scale=1.0]{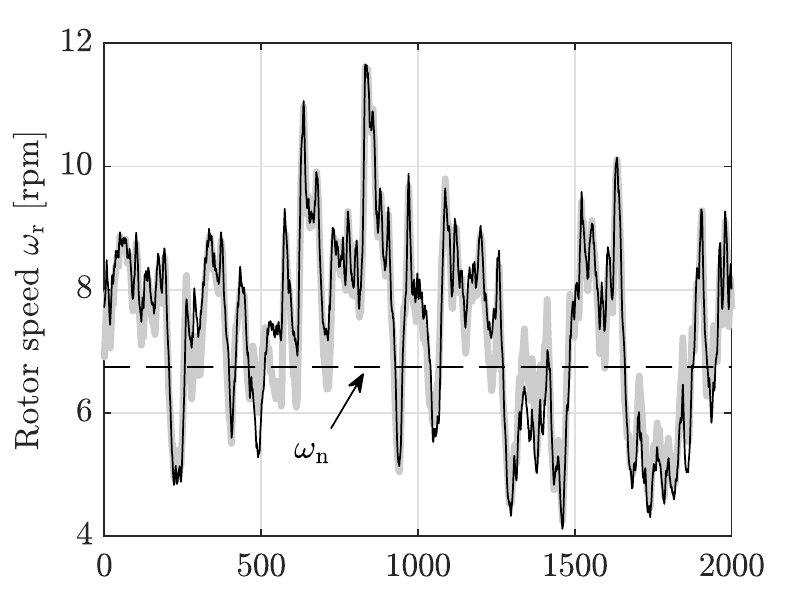}
	\includegraphics[scale=1.0]{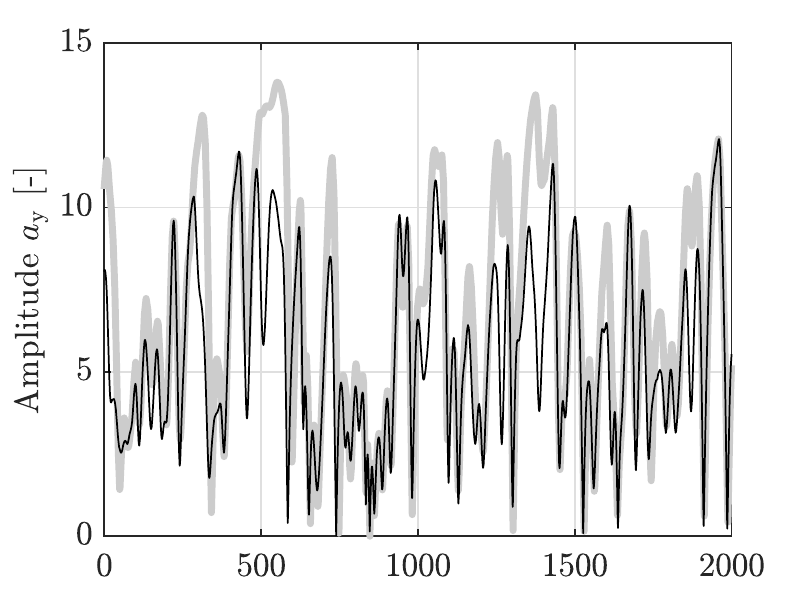}
	\includegraphics[scale=1.0]{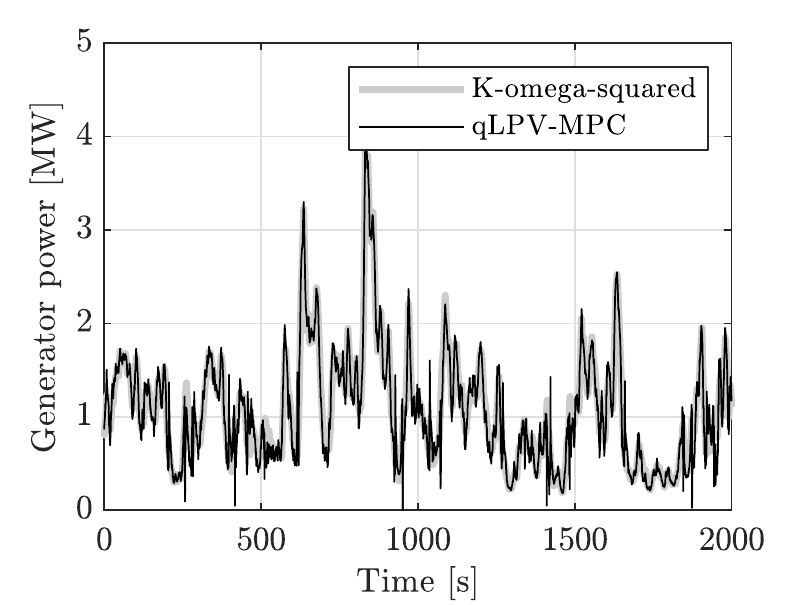}
	\includegraphics[scale=1.0]{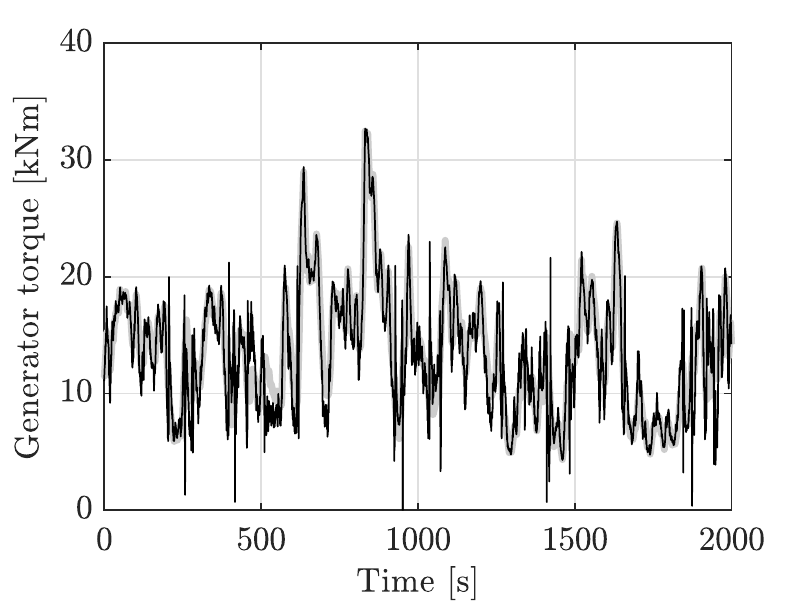}
	\caption{Simulation \textbf{Case 2} shows a comparison with conventional torque control subject to a realistic turbulent wind profile. The tower loading extremes are significantly reduced by preventing prolonged operation at the critical rotor speed. The algorithm shows to have minimal impact on the generated power.}
	\label{fig:SIM_Turbulent}
\end{figure}
\begin{figure}[t!]
	\centering
	\includegraphics[scale=1.0]{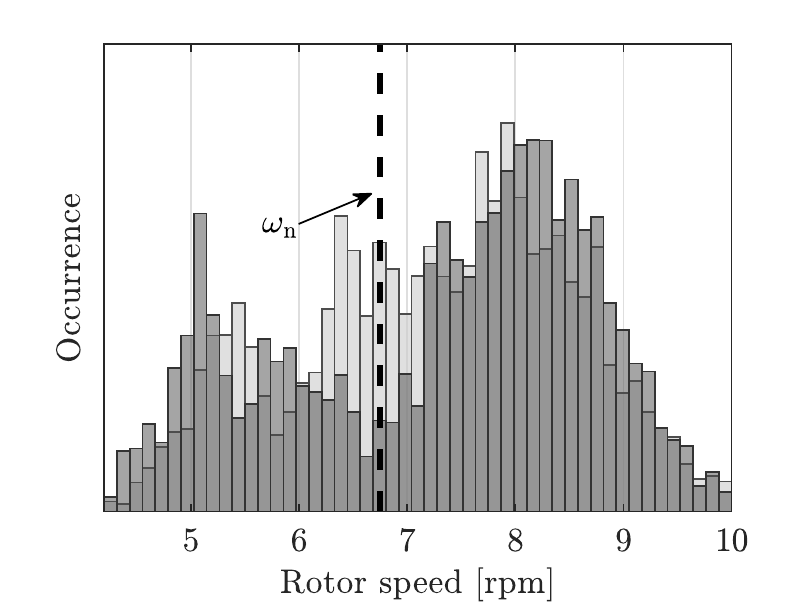}
	\includegraphics[scale=1.0]{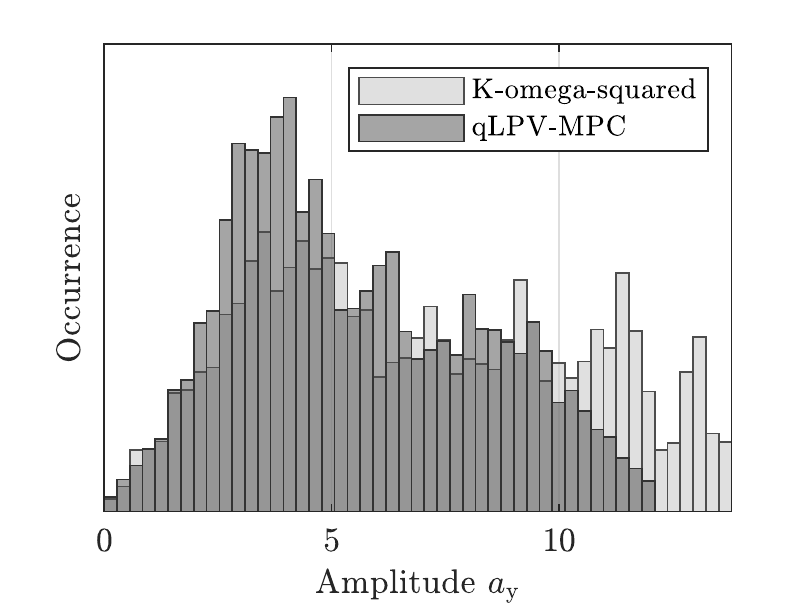}
	\caption{Histograms of the rotor speed and displacement amplitude occurrence for simulation \textbf{Case 2}. The rotor speed histogram (\textit{left}) clearly shows that the qLPV\nobreakdash-MPC algorithm prevents operation at the critical speed. Consequently, the amplitude histogram (\textit{right}) shows a reduced maximum occurrence, whereas smaller amplitudes happen more frequently, which is beneficial from a fatigue loading viewpoint.}
	\label{fig:SIM_TurbulentHistogram}
\end{figure}

Figure~\ref{fig:SIM_Slope}, presents the results for simulation \textbf{Case 1}. For this case, the in- and output weighing factors are chosen as $Q = 0.1$, $R =25$, and the prediction horizon is set to $N_\mathrm{p} = 25$. The simulation results show the ability of the algorithm to withhold the turbine from operating at a rotational speed exciting the tower natural frequency by increasing $\Delta\tau_\mathrm{g}$. Then, around $100$~s, the wind speed is sufficient for the load and power trade-off to be in favor of the latter mentioned. This is reflected by a swift reduction of the generator torque resulting in a rapid crossing of the critical rotor speed at $\omega_\mathrm{r} = \omega_\mathrm{n} = 6.75$~RPM. The tower-top displacement shows a reduction in amplitude by excitation of the natural frequency for a shorter period of time. Obviously, this comes at the expense of produced energy.

\begin{figure}[t!]
	\centering
	\subfloat[]{{\includegraphics[scale=1.075]{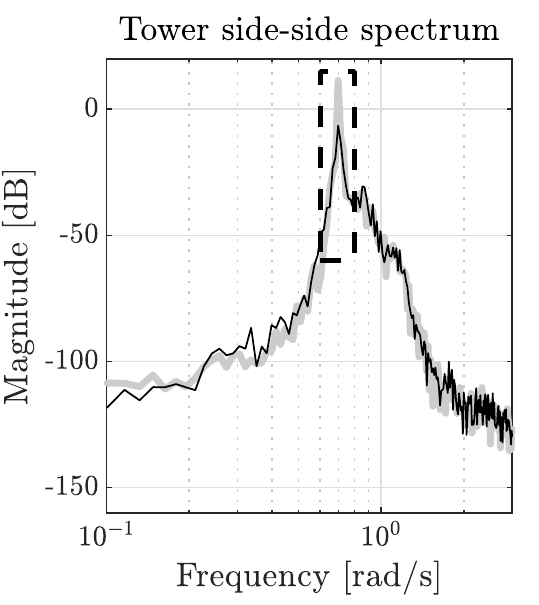} \label{fig:Tower_SideSide_Spectrum}}}
	\subfloat[]{{\includegraphics[scale=1.075]{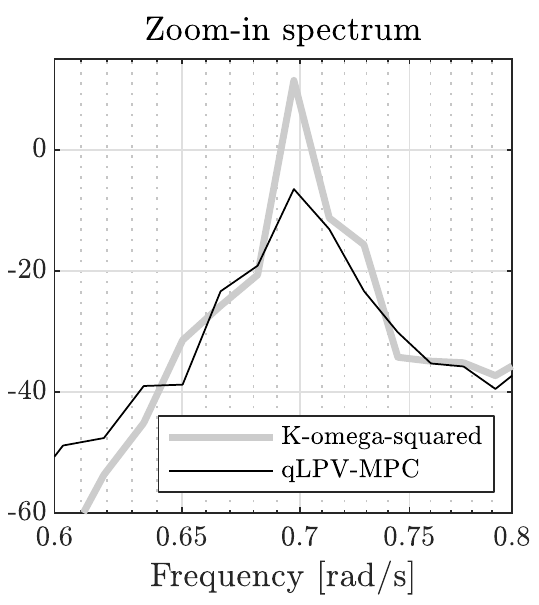}\label{fig:Tower_SideSide_SpectrumZoom}}}
	\subfloat[]{{\includegraphics[scale=1.075]{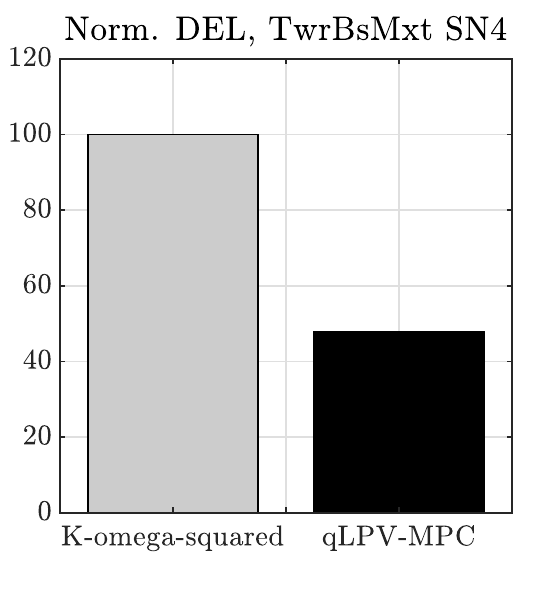} \label{fig:DELTwrBsMxt_BarPlot}}}
	\caption{\textbf{(a)} Side-side displacement spectra of the NREL~5\nobreakdash-MW turbine subject to high-fidelity turbulent wind simulations. The tower wall thickness is modified, such that a turbine side-side natural frequency of approximately $0.71$~rad\,s\textsuperscript{-1} is obtained. The blades are given a dissimilar mass to induce rotor eccentricity. Spectra for the \textit{K-omega-squared} and qLPV\nobreakdash-MPC implementations show a significant reduction of the dominant resonance. \textbf{(b)} The content in the dashed box is enlarged, and shows a peak reduction of $18$~dB. \textbf{(c)} Quantification of tower base side-side moment (TwrBsMxt) fatigue loading in terms of normalized DELs [\%], with respect to the baseline \textit{K-omega-squared} case. From a fatigue loading perspective, the qLPV\nobreakdash-MPC implementation roughly shows a significant $50$~\% DEL reduction.}
	\label{fig:Tower_SideSide}
\end{figure}

The simulation results for \textbf{Case 2} are given in Figure~\ref{fig:SIM_Turbulent}. By inspection of the rotor speed around the resonance frequency, it shows that the qLPV\nobreakdash-MPC implementation prevents operation at this speed for extended time periods. This operational strategy results in a significant decrease of tower-top displacement amplitudes. To further clarify this effect, Figure~\ref{fig:SIM_TurbulentHistogram} shows histograms of the rotor speed and displacement amplitude signals. Furthermore, Figures~\ref{fig:Tower_SideSide_Spectrum}~and~\ref{fig:Tower_SideSide_SpectrumZoom} show the sidewards displacement spectra of the two control strategies. A significant reduction of $18$~dB is attained at the turbine side-side natural frequency. Finally, a fatigue assessment is performed on the tower base side-side moment by evaluating the damage equivalent loads (DEL) from the corresponding time-domain signals using \texttt{MLife}.\cite{ref:NREL2015MLife} The DEL measure quantifies the amplitude of a certain harmonic load variation that would cause the same damage level when repeated for a given amount of cycles.\cite{ref:freebury2000DEL,ref:bossanyi2013validation} In this fatigue analysis, a W\"{o}hler-exponent of $4$ is chosen, which is a typical value for steel.\cite{ref:GL2012OffshoreGuidelines} Figure~\ref{fig:DELTwrBsMxt_BarPlot} shows the normalized DEL values with respect to the baseline case: A significant $52$~\% DEL reduction is attained. Since the analysis is based on the single load case performed in this section, the short-term DELs are calculated using the $1$~Hz equivalent load.\cite{ref:Burton2001WindEnergyHandbook} This implies that the number of cycles is equal to the simulation time, and consequently makes the analysis independent of the simulation runtime.

The generator power and torque trajectories of the control strategies show a high degree of similarity, which indicates a minimal penalty on the overall energy production. The observation is confirmed by the evaluation of the produced energy over the total simulation time, resulting in $603.34$~kWh and $601.19$~kWh for the respective baseline and qLPV\nobreakdash-MPC cases, which turns out in a negligible produced energy reduction of $0.36$~\%. The trade-off is conveniently tuned by varying the weight ratio between $Q$ and $R$.
\FloatBarrier

\section{Conclusions}\label{sec:CONCLUSIONS}
In practical scenarios, wind turbine rotors possess a mass and/or aerodynamic imbalance, which cause a periodic side-side excitation. For future turbines with higher power ratings and taller towers to be weight and cost effective, soft-soft tower configurations are being considered. Such towers are more flexible, and have their side-side natural frequency in the variable-speed domain, possibly coinciding with the rotor rotational or blade passing frequencies. 

To date, no efficient and intuitive MPC framework is available for preventing rotor speed operation at this frequency. In this paper, the dynamics of a wind turbine tower are subject to a demodulation transformation, and thereby transformed into a quasi\nobreakdash-LPV system description. The resulting qLPV model, by aggregation with a wind turbine model, is reconciled with an MPC scheme. The combination exploits the inherent properties of the qLPV model, leading to an efficient method of solving a convex optimization problem. The qLPV\nobreakdash-MPC approach involves finding the qLPV scheduling sequence by performing multiple iterative QP solves for the first time step. Subsequent time steps only require a single QP solve using a scheduling sequence warm start originating from the previous time step. By imposing an additional torque contribution, the rotor speed is prevented from operating near the tower natural frequency at the expense of reduced aerodynamic efficiency. Simulation results with artificial sloped and realistic turbulent wind profiles show that the algorithm prevents persistent excitation of the tower fundamental frequency, by sacrificing an insignificant amount of produced energy. The current work only considers the exclusion of a single excitation frequency, however, the presented framework can be extended towards the exclusion of multiple resonances.

\subsection*{Conflict of interest}
This project has been conducted in cooperation with Vestas Wind Systems A/S. 

\section*{Supporting information}

A software implementation is available as part of the online article.\cite{ref:Mulders2019DataScriptsZenodo}

\appendix
\section{-~The affine LPV model representation and discretization}\label{sec:APP_Affine}
This section presents the process of converting the LPV model derived in Section~\ref{sec:WT_Linear} to its affine form. For this, the steady-state offset values of the state, input and output values are saved for each linearization. The offsets are indicated by a $\breve{(\cdot)}$, and the following relations
\begin{align}
	\hat{\boldsymbol{q}}(t,\boldsymbol{p}^*) &= \boldsymbol{q}(t) - \breve{\boldsymbol{q}}(\boldsymbol{p}^*), \\ \hat{\boldsymbol{u}}(t,\boldsymbol{p}^*) &= \boldsymbol{u}(t) - \breve{\boldsymbol{u}}(\boldsymbol{p}^*), \\ \hat{\boldsymbol{y}}(t,\boldsymbol{p}^*) &= \boldsymbol{y}(t) - \breve{\boldsymbol{y}}(\boldsymbol{p}^*),
\end{align}
are substituted in Eq.~\eqref{eq:WT_AugmentedLinSys}, such that the affine form is obtained:
\begin{align}
	\boldsymbol{\dot{q}}(t) &= \mathbf{A}(\boldsymbol{p})(\boldsymbol{q}(t) - \breve{\boldsymbol{q}}(\boldsymbol{p}^*)) + \mathbf{B}(\boldsymbol{p})(\boldsymbol{u}(t) - \breve{\boldsymbol{u}}(\boldsymbol{p}^*)) + \boldsymbol{\dot{\breve{q}}}(\boldsymbol{p}^*)\label{eq:WT_AffineContLPVStateEq}\\
	\boldsymbol{y}(t) &= \mathbf{C}(\boldsymbol{p})(\boldsymbol{q}(t) - \breve{\boldsymbol{q}}(\boldsymbol{p}^*)) + \breve{\boldsymbol{y}}(\boldsymbol{p}^*)\label{eq:WT_AffineContLPVOutputStateEq},
\end{align}
in which $\boldsymbol{p}(t)=\boldsymbol{p}^*(t)$ indicates the current linear model in the LPV scheduling space.\cite{ref:MATLAB2018bLPV} Because a finite set of linear models is taken, the scheduling variable might fall between two model scheduling points. In this case, either the nearest offsets corresponding to the current scheduling value are taken, or a linear interpolation is performed. When the models are defined on a fine enough grid, the advantage of increased accuracy by interpolation diminishes, and therefore the \textit{nearest model} approach is employed.

As this paper uses a sample-based and fixed time-step control setup, the continuous-time system is converted to its discrete-time equivalent
\begin{align}
	\boldsymbol{q}(k+1) &= \mathbf{A}_\mathrm{d}(\boldsymbol{p}_\mathrm{k})(\boldsymbol{q}(k) - \breve{\boldsymbol{q}}(\boldsymbol{p}_\mathrm{k}^*)) + \mathbf{B}_\mathrm{d}(\boldsymbol{p}_\mathrm{k})(\boldsymbol{u}(k) - \breve{\boldsymbol{u}}(\boldsymbol{p}_\mathrm{k}^*)) + \boldsymbol{\breve{q}}(\boldsymbol{p}_\mathrm{k}^*)\label{eq:WT_AffineDiscLPVStateEq} \\
	\boldsymbol{y}(k) &= \mathbf{C}(\boldsymbol{p}_\mathrm{k})(\boldsymbol{q}(k) - \breve{\boldsymbol{q}}(\boldsymbol{p}_\mathrm{k}^*)) + \breve{\boldsymbol{y}}(\boldsymbol{p}_\mathrm{k}^*)\label{eq:WT_AffineDiscLPVOutputEq},
\end{align}
in which $k$ is the discrete time-step variable, and the matrix subscripts $(\cdot)_\mathrm{d}$ indicate the discrete time counterparts of the system and input matrices. Discretization of $\mathbf{A}$ and $\mathbf{B}$ is performed using a fourth order Runge-Kutta discretization method\cite{ref:Shampine1997Fundamentals}, of which the matrix transformation relations are given by
\begin{align}
	\mathbf{A}_\mathrm{d} &= \frac{1}{24}\mathbf{A}^{4}t_\mathrm{s}^4 + \frac{1}{6}\mathbf{A}^{3}t_\mathrm{s}^3 + \frac{1}{2}\mathbf{A}^{2}t_\mathrm{s}^2 + \mathbf{A}t_\mathrm{s} + \mathrm{I_{n}},\\
	\mathbf{B}_\mathrm{d} &= \frac{1}{24}\mathbf{A}^{3}\mathbf{B}t_\mathrm{s}^4 + \frac{1}{6}\mathbf{A}^{2}\mathbf{B}t_\mathrm{s}^3 + \frac{1}{2}\mathbf{A}\mathbf{B}t_\mathrm{s}^2 + \mathbf{B}t_\mathrm{s} + \mathrm{I_{n}}.
\end{align}
Note that the last term of Eq.~\eqref{eq:WT_AffineDiscLPVStateEq}, originating from the left-hand side of the equation, is in the discrete-time case taken at the current time instant, as the output for scheduling the next state offset $\boldsymbol{\breve{q}}$ is unavailable at time step $k$.

\section{-~LPV forward propagation matrices}\label{sec:APP_ForwardPropagation}
This section defines the LPV forward-propagation matrices of Eq.~\eqref{eq:WT_MPC_Equation}:
\begin{align}
	\boldsymbol{Y}_{k+1} &= \begin{bmatrix} 
		y_{k+1}\\
		y_{k+2}\\
		\vdots\\
		y_{k+N_p}
	\end{bmatrix},
	~
	\boldsymbol{\breve{Y}}_{k+1}(\mathbf{P}_{k}) = \begin{bmatrix} 
		\breve{y}(\boldsymbol{p}_\mathrm{k+1})\\
		\breve{y}(\boldsymbol{p}_\mathrm{k+2})\\
		\vdots\\
		\breve{y}(\boldsymbol{p}_\mathrm{k+N_p})
	\end{bmatrix},
	~
	\Delta\boldsymbol{U}_{k}(\mathbf{P}_{k}) = \begin{bmatrix}
		u_{k} - \breve{u}(\boldsymbol{p}_\mathrm{k})\\
		u_{k+1} - \breve{u}(\boldsymbol{p}_\mathrm{k+1})\\
		\vdots\\
		u_{k+\mathrm{N_p}-1} - \breve{u}(\boldsymbol{p}_\mathrm{k+\mathrm{N_p}-1})
	\end{bmatrix}\nonumber\\
	\mathbf{H}(\mathbf{P}_{k}) &= \begin{bmatrix}
		C(\boldsymbol{p}_{k+1})A(\boldsymbol{p}_{k})\\
		C(\boldsymbol{p}_{k+2})A(\boldsymbol{p}_{k+1})A(\boldsymbol{p}_{k})\\
		\vdots\\
		C(\boldsymbol{p}_{\mathrm{k+N_p}})A(\boldsymbol{p}_{k+\mathrm{N_p}-1})\dots A(\boldsymbol{p}_{k})
	\end{bmatrix},\quad
	\mathbf{D}(\mathbf{P}_{k}) = \begin{bmatrix}
		&D(\boldsymbol{p}_{k+1})&0&\cdots&0 \\
		&0&D(\boldsymbol{p}_{k+2})&\cdots&0\\
		&\vdots&&\ddots&\vdots\\
		&0&\cdots&&D(\boldsymbol{p}_{k+N_p})
	\end{bmatrix},\nonumber\\
	\mathbf{S}(\mathbf{P}_{k}) &= \begin{bmatrix}
		&C(\boldsymbol{p}_{k+1})B(\boldsymbol{p}_{k}) &0 &\dots &0\\
		&C(\boldsymbol{p}_{k+2})A(\boldsymbol{p}_{k+1})B(\boldsymbol{p}_{k}) &C(\boldsymbol{p}_{k+2})B(\boldsymbol{p}_{k+1}) &\dots &0 \\
		&\vdots & &\ddots &\vdots\\
		&C(\boldsymbol{p}_\mathrm{k+N_p})A(\boldsymbol{p}_{\mathrm{k+N_p}-1})\dots A(\boldsymbol{p}_{k+1})B(\boldsymbol{p}_{k}) &\cdots&& C(\boldsymbol{p}_{\mathrm{k+N_p}})B(\boldsymbol{p}_{\mathrm{k+N_p}-1})
	\end{bmatrix},\nonumber\\
	\mathbf{L}(\mathbf{P}_{k}) &= \begin{bmatrix}
		&C(\boldsymbol{p}_{k+1}) &0 &\dots &0\\
		&C(\boldsymbol{p}_{k+2})A(\boldsymbol{p}_{k+1}) &C(\boldsymbol{p}_{k+2}) &\dots &0 \\
		&\vdots & &\ddots &\vdots\\
		&C(\boldsymbol{p}_{\mathrm{k+N_p}})A(\boldsymbol{p}_{k+\mathrm{N_p}-1})\dots A(\boldsymbol{p}_{k+1}) &\cdots&& C(\boldsymbol{p}_{\mathrm{k+N_p}})
	\end{bmatrix},\quad
	\Delta\boldsymbol{\breve{X}}_{k}(\mathbf{P}_{k}) = \begin{bmatrix}
		\breve{x}(\boldsymbol{p}_\mathrm{k}) - \breve{x}(\boldsymbol{p}_\mathrm{k+1})\\
		\breve{x}(\boldsymbol{p}_\mathrm{k+1}) - \breve{x}(\boldsymbol{p}_\mathrm{k+2})\\
		\vdots\\
		\breve{x}(\boldsymbol{p}_\mathrm{k+N_p-1}) - \breve{x}(\boldsymbol{p}_\mathrm{k+N_p})
	\end{bmatrix},\nonumber
\end{align}


\bibliography{TowerVibrationMPC}%

\end{document}